\documentclass[twocolumn,alpha-refs,serif]{wiley-article}

\usepackage{orcidlink}
\usepackage{siunitx}
\usepackage{cleveref}

\usepackage{blindtext}
\usepackage{booktabs}
\usepackage{graphbox}
\usepackage{multirow}

\papertype{Research Article}

\title{A machine-learning approach to thunderstorm forecasting through post-processing of simulation data}

\author[1]{Kianusch Vahid Yousefnia\,\orcidlink{0000-0003-2644-2539}}
\author[1]{Tobias Bölle\, \orcidlink{0000-0003-3714-6882}}
\author[1]{Isabella Zöbisch\, \orcidlink{0000-0003-2035-7931}}
\author[1]{Thomas Gerz\, \orcidlink{0000-0003-2923-3126}}

\affil[1]{Deutsches Zentrum für Luft- und Raumfahrt (DLR), Institut für Physik der Atmosphäre, Oberpfaffenhofen, Germany}

\corraddress{Kianusch Vahid Yousefnia, DLR Oberpfaffenhofen, Institut für Physik der Atmosphäre, Münchner Str. 20, D-82234 Wessling, Germany}
\corremail{kianusch.vahidyousefnia@dlr.de}

\fundinginfo{Internal DLR project DIAL}

\runningauthor{Vahid Yousefnia et al.}

\begin{document}

\begin{frontmatter}
\maketitle

\begin{abstract}
Thunderstorms pose a major hazard to society and economy, which calls for reliable thunderstorm forecasts.
In this work, we introduce a Signature-based Approach of identifying Lightning Activity using MAchine learning (SALAMA), a feedforward neural network model for identifying thunderstorm occurrence in numerical weather prediction (NWP) data. The model is trained on convection-resolving ensemble forecasts over Central Europe and lightning observations. Given only a set of pixel-wise input parameters that are extracted from NWP data and related to thunderstorm development, SALAMA infers the probability of thunderstorm occurrence in a reliably calibrated manner.
For lead times up to eleven hours, we find a forecast skill superior to classification based only on NWP reflectivity. Varying the spatiotemporal criteria by which we associate lightning observations with NWP data, we show that the time scale for skillful thunderstorm predictions increases linearly with the spatial scale of the forecast.

\end{abstract}
\end{frontmatter}

\section{Introduction}
While thunderstorms undoubtedly constitute inspiring natural spectacles that move any human being to a certain extent, their impact in the form of lightning, strong winds and heavy precipitation (including hail) is hazardous to society and economy. Besides the small but real chance of being struck by lightning \citep{Holle2016}, thunderstorms pose a threat to crops  and lifestock \citep{Holle2014} as well, and are known to trigger wild fires \citep{Veraverbeke2017}. In addition, they constitute a major safety concern for aviation \citep{Gerz2012, Borsky2019}. Furthermore, thunderstorms and lightning damage electrical infrastructure such as wind turbines \citep{Yasuda2012}, which jeopardizes the transition to sustainable energy production. Finally, since the number of severe thunderstorms is expected to increase due to climate change \citep{Diffenbaugh2013, Raedler2019}, accurate thunderstorm forecasts become ever more relevant.

Thunderstorm forecasts with lead times of more than one hour usually rely on numerical weather prediction (NWP).
This method consists of simulating the future atmospheric state by numerically solving equations derived from the laws of physics. The accuracy of NWP has improved with the advent of high-performance computing, the increased availability of observational data through satellite imagery, as well as advances in data assimilation \citep{Bauer2015, Yano2018}. In order to use NWP data for thunderstorm predictions, one needs to know how thunderstorms manifest themselves in terms of the NWP output fields. In a post-processing step, this knowledge is then used to identify signs of thunderstorm occurrence in simulation data.


Various ideas for identifying signs of thunderstorm occurrence have been put forward in recent years. For instance, post-processing of NWP data has been blended with nowcasting methods \citep{Kober2012, Hwang2015}. Empirical knowledge on convective activity has been translated into expert systems using fuzzy logic \citep{Lin2012, Li2021}. The fuzzy logic technique allows the construction of decision rules for thunderstorm occurrence based on domain knowledge.  Lately, machine learning (ML) methods based on artificial neural networks have gained popularity. These methods generalize the fuzzy logic approach in the sense that decision rules are constructed by solving a data-driven optimization problem. Previous studies include neural networks with relatively few neurons \citep{Ukkonen2019, Kamangir2020, Sobash2020, Jardines2021}, as well as deep neural networks with convolutional layers and millions of trainable parameters \citep{Geng2021, Zhou2022}. Findings suggest that neural network models are more skillful at predicting thunderstorm occurrence than comparable ML approaches like random forests \citep{Herman2018, Ukkonen2019}.
In order to learn predicting thunderstorm occurrence, supervised ML methods require a ground truth of thunderstorm activity. It may be provided by satellite imagery \citep{Jardines2021, Zhou2022}, radar data \citep{Gagne2017, Burke2020, Leinonen2022}, storm reports \citep{Loken2020, Sobash2020}, and lightning \citep{Ukkonen2019, Geng2021}.

The promising results in ML have encouraged us to apply neural network methods to historical simulation data of the ICOsahedral Nonhydrostatic D2 Ensemble Prediction System (ICON-D2-EPS), an NWP ensemble model for Central Europe with a horizontal resolution of  ca. \SI{2}{\kilo\meter} \citep{Zaengl2015, Reinert2020}. ICON-D2-EPS is a limited-area model which explicitly resolves convection and is run operationally by the German Meteorological Service (DWD). To the best of our knowledge, neural networks have not yet been employed for the identification of thunderstorm occurrence in ensemble data with a comparable horizontal resolution. In this work, we present the neural network model SALAMA (Signature-based Approach of identifying Lightning Activity using MAchine learning). It has been trained to predict thunderstorm occurrence through the post-processing of  simulation data. 

In \cref{sec:data}, we describe how independent datasets for the training, testing and validation of our model have been compiled from NWP forecasts and lightning data. Details on the ML architecture are provided in \cref{sec:methods}. 
While thunderstorm occurrence is identified in a pixel-wise manner, we systematically vary the spatiotemporal criteria by which the lightning observations are associated with the NWP data. This enables us to study the effect of different spatial scales on the model identification skill and allows us to estimate the advection speed of thunderstorms.
Further results are presented in \cref{sec:results} and demonstrate that, for lead times up to at least eleven hours, SALAMA is more skillful than a baseline method based only on NWP reflectivity. In addition, we show a linear relationship between the spatial resolution scale of our model and the time scale during which skill decreases with lead time. This is consistent with earlier findings that resolving smaller scales brings faster growing forecast errors about \citep{Lorenz1969,Selz2015}.

\section{Data}\label{sec:data}
We  collected simulation data from the ICON-D2-EPS ensemble model, as well as lightning observations from the lightning detection network LINET \citep[LIghtning detection NETwork,][]{Betz2009}.
The simulations were used to extract predictors of thunderstorm occurrence, while lightning observations serve as ground truth.

\subsection{Study region and period}
The model domain of ICON-D2-EPS covers the areas of Germany, Switzerland, Austria, Denmark, Belgium, the Netherlands and parts of the neighboring countries. For our study, we cropped the model domain at its borders by approximately \SI{100}{\kilo\meter}  to reduce boundary computation errors. In a cylindrical projection, our study region corresponds to a rectangle with the southwest corner located at \ang{45}N, \ang{1}E, the northeast corner located at \ang{56}N, \ang{16}E and all sides being either parallels or meridians (\cref{fig:case_visualization}).

There are daily model runs every three hours starting at 00~UTC. We collected simulation data from June to August 2021 over the entire study region in hourly steps, taking always the latest available forecast for each hour. Following this procedure results in forecasts with lead times of \SI{0}{\hour}, \SI{1}{\hour} or \SI{2}{\hour}.

Each model run has 20 ensemble members which differ from each other in a manner consistent with the NWP uncertainty in the initial conditions, model error, and boundary conditions \citep{Reinert2020}. In \cref{sec:lead-dependence}, we will relate NWP forecast uncertainty, estimated by ensemble variability, to ML model skill.

\subsection{NWP predictors}


The atmospheric fields used as predictors of thunderstorm occurrence in this study are given in \cref{tab:input_features}. They have been selected as follows: We considered as candidate predictors all two-dimensional fields provided in ICON-D2-EPS, as well as two ICON-D2-EPS pressure-level fields associated with deep moist convection in the literature, namely the relative humidity at \SI{700}{\hecto\pascal} and the vertical wind speed in pressure coordinates at \SI{500}{\hecto\pascal} \citep{Li2021}. In addition, we stipulated that the predictors be available on the open-data server of the DWD (https://opendata.dwd.de, last visit: 2023-03-14), such that the trained model can eventually be used in real-time.
For a given candidate input field, 
we compared histograms of the field value distribution during and in the absence of thunderstorm occurrence and kept only fields that differed significantly in the two distributions.  

As shown in \cref{tab:input_features}, all predictors can be related to thunderstorm activity through physical mechanisms like instability and moisture. In particular, our selection process has led to predictors that agree with findings in the literature \citep{Ukkonen2019, Jardines2021, Leinonen2022}. Conversely, convective inhibition (CIN), which is sometimes listed as a convective predictor \citep{Kamangir2020}, has not passed the selection process. This is likely due to the fact that we have checked for predictive power in terms of developed thunderstorms. CIN, however, correlates with the hours leading up to a thunderstorm and has been removed once the storm reaches its mature stage.

It is worth stressing that we have excluded certain parameters on purpose, namely the geographical location of a thunderstorm event, the time of the day, and the time of the year. In doing so, we assume the existence of a universal signature shared by all thunderstorms, irrespectively of where and when they occur. In addition, the list of predictors does not include the lead time of the forecast. We check in \cref{sec:results} whether our model, which has been trained on data with lead times between \SI{0}{\hour} and \SI{2}{\hour}, displays skill on data with longer lead times.

\begin{table*}[htbp]
    \centering
    \caption{List of the 21 input parameters used in the study ("DIA": including sub-grid scale). }
    \label{tab:input_features}
    \begin{tabular}{lll}
    \toprule
  physical significance & ICON parameter name  & description \\
   \midrule
        instability & CAPE\textunderscore ML & mixed-layer convective available potential energy\\
                 & CEILING &  ceiling height\\
                 & OMEGA500 & vertical wind speed in pressure coordinates at \SI{500}{\hecto\pascal}\\
                & PS & surface pressure \\
                 & PMSL & surface pressure reduced to mean sea level \\
        cloud cover & CLCH & high level clouds (0-\SI{400}{\hecto\pascal}) \\
            & CLCM & mid-level clouds (400-\SI{800}{\hecto\pascal}) \\
            & CLCL & low-level clouds (\SI{800}{\hecto\pascal} to soil) \\
            & CLCT & total cloud cover \\
        precipitation and & DBZ\textunderscore CMAX & maximal radar reflectivity \\
        moisture& ECHOTOP & echotop pressure \\
        & RELHUM700 & relative humidity at \SI{700}{\hecto\pascal} \\
        & RELHUM\textunderscore 2M & \SI{2}{\meter} relative humidity \\
    column-integrated & TQC, TQC\textunderscore DIA & cloud water \\
       water quantities & TQG & graupel \\
        & TQI, TQI\textunderscore DIA & ice \\
        & TQV, TQV\textunderscore DIA & water vapor \\
        & TWATER & total water content \\
        \bottomrule
    \end{tabular}
\end{table*}

\subsection{Lightning observations}\label{sec:lightning}
In supervised learning, ML models are trained on  data for which the ground truth is known. For this reason, we required knowledge of thunderstorm occurrence for our study domain and period. By reason of their high detection efficiency and spatial accuracy over the entire study region, we employed lightning observations to assess the occurrence of thunderstorms.
Specifically, we resorted to the LINET network \citep{Betz2009}, which exploits the radio spectrum to continuously measure strokes of lightning over Europe. The technology achieves a detection efficiency of more than \SI{95}{\percent} and an average location accuracy of \SI{150}{\meter}. While the technology is able to differentiate between cloud-to-ground and intracloud flashes, we have considered all lightning events as we are only interested in the yes/no occurrence of thunderstorm activity.

Given a set of predictors retrieved from a grid point $\mathbf{x}$ on the study domain at time $t$ during the study period, we considered thunderstorm activity to occur at $(\mathbf{x},t)$ if a flash was detected at any $(\mathbf{x}_l, t_l)$ with
\begin{equation}
    \left\| \mathbf{x}-\mathbf{x}_l \right\| < \Delta r, \quad \text \quad |t-t_l| < \Delta t,
\end{equation}
where $\left\|\cdot\right\|$ denotes the great-circle distance between $\mathbf{x}$ and $\mathbf{x}_l$. We trained our model with different values for the spatial and temporal thresholds $\Delta r$ and $\Delta t$ in order to study the relationship between them and classification skill systematically.

\subsection{Compiling independent data sets}\label{sec:datasets}
The data obtained from NWP and lightning observations can be considered a set of tuples $(\bm{\xi}, y)$, where $\bm{\xi}\in \mathbb{R}^n$ denotes the $n=21$ input parameters and $y \in \{0,1\}$ corresponds to a label of the ground truth (1: thunderstorm occurrence, 0: no thunderstorm occurrence).  As the input fields were provided on a triangular grid, we first performed an interpolation onto a $\ang{0.125}\times\ang{0.125}$ longitude/latitude grid. The labels were produced on the same grid. For each full hour during the study period, for each ensemble member and for each grid point, we fetched the corresponding tuple $(\bm{\xi}, y)$, taking always the latest available forecast. 

\begin{figure}[htbp]
\centering
\includegraphics[width=\columnwidth]{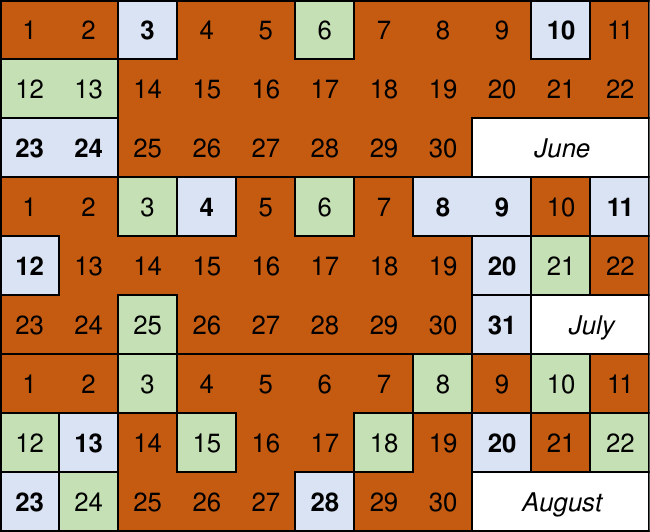}
\caption{Days (from 8~UTC to 8~UTC) during the summer of 2021 which were used for compiling the datasets for training (dark brown), testing (light blue with bold numerals) and validation (light green). The days have been distributed at random among the three sets.}
\label{fig:training_days}
\end{figure}

The resulting large collection of tuples was then divided into three statistically independent data sets: The training set is used only for training the neural network model (a precise definition of training is given in \cref{sec:model_description}), while its skill is measured on a test set with data that the model has not seen during training. A third data set, the validation set, is used to monitor training progress (\cref{sec:model_description}). 
In an attempt to assure statistical independence between the data sets, we took two measures. First, assuming possible day-to-day correlations in the input parameters (e.g. induced by the synoptic scale) to be negligible for convective events with life spans of the order of a few hours, we used separate days for training, testing and validation. 
 In addition, we took into account that intense thunderstorms that form in the afternoon may well live on after 0~UTC. We therefore defined days to begin at 8~UTC, a time of the day chosen by checking when lightning activity in the collected data is minimal. The latter measure prevents data from one thunderstorm at different times to appear in separate data sets. 
\Cref{fig:training_days} offers an overview of the days contained in each data set. The days were randomly distributed among the three sets. Additionally, we randomly subsampled the data such that the training set consists of \num{4e5} tuples, and the test and validation sets each contain \num{e5} tuples.

The rarity of thunderstorms makes predicting their occurrence more challenging as ML models tend to struggle with learning from unbalanced datasets \citep{Sun2009}. As a matter of fact, we verified that when trained on a climatologically consistent dataset, our model would predict the majority class (i.e. no thunderstorm) at every occasion. We therefore undersampled the majority class in the training set, such that both labels appear equally frequently (class balance). On the other hand, the validation and testing set remain climatologically consistent since we wish to quantify model performance in a realistic setting. Having different sample climatologies in the training and test set, however, requires model output calibration, which is discussed in \cref{sec:model_reliability}.

\section{Methods}\label{sec:methods}
In this section, we provide details on SALAMA, focusing on how it has been trained and calibrated. In addition, we introduce metrics for the evaluation of model skill and present a baseline model for comparison.

\subsection{Model description} \label{sec:model_description}
It is worthwhile to introduce some ML terminology. The three data sets used for training, testing and validation (\cref{sec:datasets}) are made up of examples $(\bm{\xi}, y)$. Each example consists of a pattern $\bm{\xi} \in \mathbb{R}^n$ of $n$ input features and a label $y\in\{ 0,1\}$.

\begin{figure}[htbp]
\centering
\includegraphics[width=\columnwidth]{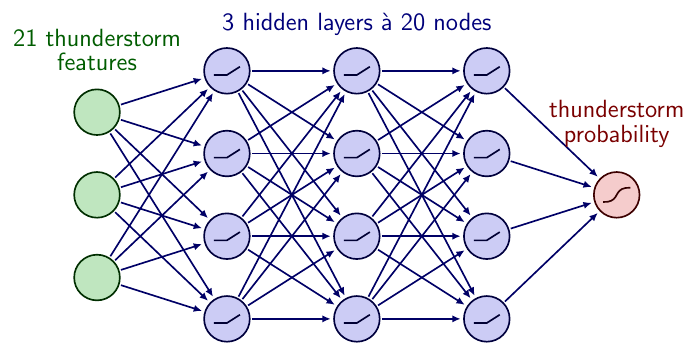}
\caption{(Color online) The architecture of SALAMA: Input features are scaled to order 1. We use rectified linear units as activation functions in the hidden layers. A sigmoid function maps the output layer  to the open interval $(0,1)$.}
\label{fig:architecture}
\end{figure}

Given a pattern $\bm{\xi}$, the problem at hand is to infer the probability of thunderstorm occurrence, which constitutes a task known as binary classification. In the following, we consider both the pattern and its corresponding label to originate from a random experiment. Therefore, let $\mathbf{\Xi}$ be an $n$-dimensional  random variable for the pattern and let $Y$ be a random variable of thunderstorm occurrence (1: thunderstorm, 0: no thunderstorm). We are interested in $P(Y=1 | \mathbf{\Xi}=\bm{\xi})$, namely the conditional probability of thunderstorm occurrence if the pattern is known. A feedforward artificial neural network model is a function $f: \mathbb{R}^n\to (0,1)$ that models the relationship between the input pattern and the corresponding probability of thunderstorm occurrence. We refer to $f$ simply as neural network. Neural networks use compositions of matrix multiplications, as well as non-linear operations referred to as activation functions. The architecture of our neural network is presented in \cref{fig:architecture}: It consists of the input and output layer as well as hidden layers, where each layer is a vector of numbers obtained from the previous layer by one matrix multiplication and by applying an activation function to the result in a component-wise manner. The complexity of $f$ is adjustable through the number of hidden layers and the size of each hidden layer, i.e. the number of nodes. Our model has three hidden layers and 20 nodes per hidden layer. Moreover, we use rectified linear units for the hidden layers and a sigmoid function to map the output layer to a probability between zero and one. 

The entries, also referred to as weights, of the matrices that connect the layers are adjusted according to the data in the training set. We therefore add a subscript $\mathbf{w}\in\mathbb{R}^d$ to $f$  to express the dependence on the $d$ weights. If $f_\mathbf{w}$ constitutes an accurate representation of the conditional probability of thunderstorm occurrence, i.e. $f_\mathbf{w}(\bm{\xi}) = P(Y=1 | \mathbf{\Xi}=\bm{\xi})$, then the likelihood of observing a label $y$ for a given input feature $\bm{\xi}$ reads
\begin{equation}
L(\mathbf{w} | \bm{\xi},y) = 
\begin{cases}
f_\mathbf{w}(\bm{\xi}), & y =1 \\
1-f_\mathbf{w}(\bm{\xi}), & y=0 
\end{cases}
\end{equation}

Denote by $(\bm{\xi}^{(i)},y^{(i)})_{i=1\dots N}$ the training set with $N$ examples. The most likely configuration of weights, given the training set, is then obtained by minimizing the negative logarithm of the likelihood function,
\begin{equation}
-\log \mathcal{L} (\mathbf{w}) = -\sum_{i=1}^{N} \log L\left(\mathbf{w} | \bm{\xi}^{(i)},y^{(i)}\right),
\label{eq:loglikelihoodloss}
\end{equation}
with respect to the weights.
The expression in \cref{eq:loglikelihoodloss} is referred to as binary cross-entropy loss function in ML terminology. The process of determining the weights that minimize loss is called training.  We trained SALAMA using the robust iterative stochastic method Adam \citep{Kingma2014}. However, if one used the configuration of weights which minimizes \cref{eq:loglikelihoodloss} exactly, a neural network would likely suffer from overfitting, i.e. learning parts of the noise in the data as well. To this end, we implemented an early stopping procedure, in which loss was monitored on the validation set during training. Once the validation loss no longer decreased, training was stopped. 

Before training, each input feature has been scaled in a way that its sample standard deviation in the training set is of the order of unity. In addition, we trained not only on the architecture presented in \cref{fig:architecture} but also varied the number of hidden layers,  as well as the number of nodes per layer. We found that once a certain complexity was reached in terms of the size of the network, adding new nodes or layers had no effect on the validation loss at the end of training. The architecture in \cref{fig:architecture} constitutes the smallest network for which this complexity threshold has been exceeded.

A conceptional issue we would like to address concerns our use of ensemble data. In our data sets, a given $\bm{\xi}^{(i)}$ has been retrieved from a specific (but arbitrary) member of the NWP ensemble. Consequently, $f_\mathbf{w}(\bm{\xi})$ refers to the probability of thunderstorm occurrence associated with this individual member. With the exception of a study of ensemble spread in \cref{sec:lead-dependence}, the results in the remainder of this work are, therefore, based on probabilities for one individual member.

\subsection{Analytic model calibration}\label{sec:model_reliability}

In order to address the climatological rarity of thunderstorm occurrence, we have artificially increased the fraction of positive examples in the data set used for the training of our neural network (\cref{sec:datasets}). In this section, we explain why this dataset modification causes our model to be miscalibrated, and derive an analytic correction for model output calibration.

It is crucial to understand that if the trained model were naively applied to a test set with a different fraction of positive examples than in the training set, the produced probabilities would be inconsistent with the observed relative frequency of thunderstorm occurrence.
In order to see this, we use Bayes' theorem to expand the conditional probability of thunderstorm occurrence given a pattern $\bm{\xi}$, which yields:
\begin{equation}
P(Y=1 | \mathbf{\Xi}=\bm{\xi})  = \frac{P(\mathbf{\Xi}=\bm{\xi} | Y = 1) P(Y=1)}{P(\mathbf{\Xi}=\bm{\xi})}.
\end{equation}
The denominator can be expressed as
\begin{equation}
\begin{split}
P(\mathbf{\Xi}=\bm{\xi}) &= P(\mathbf{\Xi}=\bm{\xi} | Y = 1)P(Y=1)  \\ 
&+  P(\mathbf{\Xi}=\bm{\xi} | Y = 0)P(Y=0).
\end{split}
\end{equation}
Let $P(Y=1)=1-P(Y=0) = g$, where $g$ denotes the climatological probability of thunderstorm occurrence with no prior knowledge. Then,
\begin{equation}
P(Y=1 | \mathbf{\Xi}=\bm{\xi})  = \frac{1}{1+(1-g)R(\bm{\xi})/g} \label{eq:g-dep},
\end{equation}
where the residual function $R(\bm{\xi})=P(\mathbf{\Xi}=\bm{\xi} | Y = 0)/P(\mathbf{\Xi}=\bm{\xi} | Y = 1)$ is not expected to depend on $g$. Nevertheless, \cref{eq:g-dep} shows that the conditional probability of thunderstorm occurrence does carry an implicit $g$-dependence through the prefactor $(1-g)/g$. The training set contains an increased fraction $\tilde{g}$ of positive examples (in our work: $\tilde{g}=1/2$), while the corresponding fraction in the test set is (up to fluctuations due to the finite sample size) equal to the climatological value $g$. During training, the neural network, therefore, learns to produce the following model output
\begin{equation}
f_\mathbf{w}(\bm{\xi},\tilde{g})  = \frac{1}{1+(1-\tilde{g})R(\bm{\xi})/\tilde{g}}.
 \label{eq:g-dep2}
\end{equation}
When we want to apply our neural network to a dataset with $g\neq \tilde{g}$, the correct probability output reads
\begin{equation}
f_\mathbf{w}(\bm{\xi}, g) = \frac{f_\mathbf{w}(\bm{\xi},\tilde{g})}{f_\mathbf{w}(\bm{\xi},\tilde{g}) + \frac{1-g}{g}\frac{\tilde{g}}{1-\tilde{g}}(1-f_\mathbf{w}(\bm{\xi},\tilde{g}))},
\label{eq:probability_test}
\end{equation}
which can be derived by solving \cref{eq:g-dep2} for $R(\bm{\xi})$ and substituting the result into \cref{eq:g-dep}. On the other hand, if the sample climatology of training set and test set are equal ($\tilde{g}=g$), \cref{eq:probability_test} yields $f_\mathbf{w}(\bm{\xi}, g) = f_\mathbf{w}(\bm{\xi}, \tilde{g})$, i.e. no probability correction is needed.

\begin{figure}[htbp]
\centering
\includegraphics[width=\columnwidth]{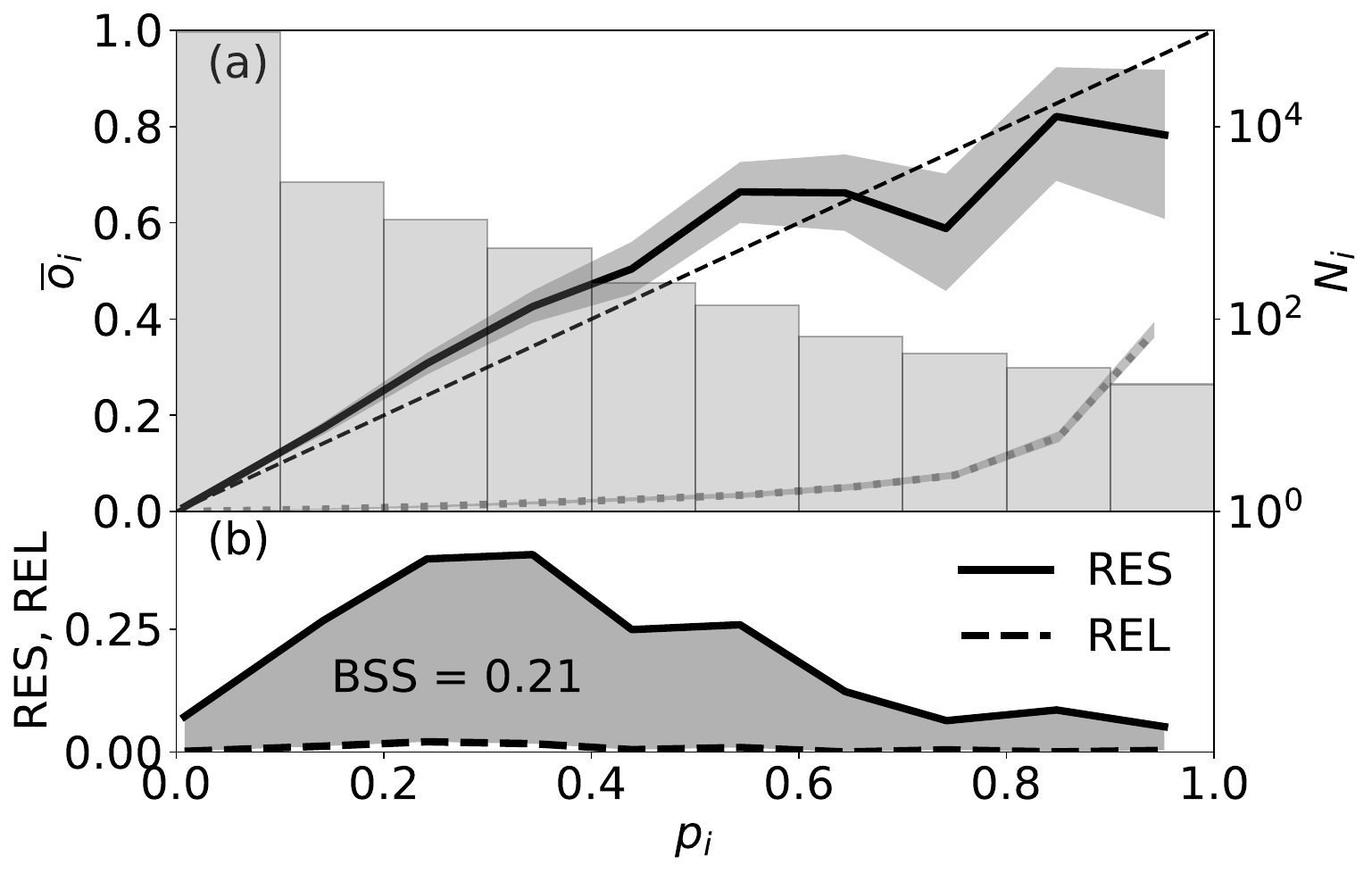}
\caption{
Reliability diagram of SALAMA, evaluated for the test set with the label configuration $\Delta r = \SI{15}{\kilo\meter}, \Delta t = \SI{30}{\minute}$ (\cref{sec:lightning}). (a) Calibration curve after applying probability correction \eqref{eq:probability_test} (black solid line), and before (grey light dotted line), and histogram of examples per bin. Perfect reliability is indicated by a dashed diagonal. Shaded band corresponds to the symmetric \SI{90}{\percent} confidence interval obtained by \num{200} bootstrap resamples. (b) Bin-wise resolution and reliability (\crefrange{eq:res}{eq:rel}) and their relation to the Brier skill score (BSS, \cref{sec:skill_scores}) as a function of model probability.
}
\label{fig:reliability_diagram}
\end{figure}

If  the model probability output is consistent with the observed relative frequency of thunderstorm occurrence, the model forecasts are referred to as reliable.
In order to check whether our neural network  provides reliable forecasts, we used the test set to produce a reliability diagram. For this purpose, one partitions the interval $(0,1)$ of possible forecast probabilities into bins. For each bin, one considers all examples whose model probability falls into the bin. Then, one computes the relative frequency of thunderstorm occurrence and plots it against the bin-averaged model probability per bin. The resulting curve is referred to as calibration function. An example for one configuration of lightning labels is shown in \cref{fig:reliability_diagram} (a), for which 10 equidistant bins have been used. 
Shown are two calibration functions: The light grey line corresponds to a calibration function \emph{without} any probability correction, while the solid black line results from applying \eqref{eq:probability_test} to our model output. 
The uncertainty on the observed frequency spans the 5th and 95th percentiles of fluctuations and has been estimated through a bootstrap resampling procedure similar to \cite{Broecker2007}: By drawing with replacement, one produces variations of the original test set and considers the sample-to-sample fluctuations of observed relative frequencies. 

The uncalibrated line severely overestimates the relative frequency of thunderstorm occurrence at all model probabilities. As has been worked out, this is not a result of faulty training but stems from having different sample climatologies in the training and test sets. After calibration, however, our model produces reliable forecasts for probabilities close to 0 and 1. On the other hand, our model slightly underestimates the relative frequency of thunderstorm occurrence for forecast probabilities below \num{0.6}. Further calibration could be done using statistical methods like isotonic regression \citep{Niculescu-Mizil2005}, which is beyond the scope of this work. Instead, we consider our model sufficiently reliable and appreciate that the level of reliability has been attained by means of the analytical correction \eqref{eq:probability_test} alone.

In addition to calibration curves, binning the forecast probabilities allows the introduction of two useful metrics of classification skill. Of the $N$ examples in the test set, let $N_i$ fall into bin $i$ with bin width $\Delta p_i$, bin-averaged model probability $p_i$ and observed relative frequency $\overline{o}_i$ of thunderstorm occurrence. We then define the following two bin-wise terms:
\begin{align}
\text{RES}_i &=  \frac{1/\Delta p_i}{g(1-g)}\frac{N_i}{N} (p_i - g)^2 \label{eq:res}\\
\text{REL}_i &=  \frac{1/\Delta p_i}{g(1-g)}\frac{N_i}{N} (p_i - \overline{o}_i)^2 \label{eq:rel}
\end{align}
Up to a factor $g(1-g)$ known as uncertainty term, the sums $\sum_i \Delta p_i \text{RES}_i$ and $\sum_i \Delta p_i \text{REL}_i$ are called the resolution and reliability, respectively, of the model. Resolution measures forecast variance, with higher values of resolution indicating a better ability of the model to differentiate between thunderstorm and non-thunderstorm patterns \citep{Toth2003}. Reliability quantifies the mean squared deviation of the calibration curve from the diagonal. The bin-wise terms defined in \crefrange{eq:res}{eq:rel} and shown in \cref{fig:reliability_diagram} (b) offer an overview of how much each probability bin contributes to reliability and resolution. For instance, resolution is most impacted by examples with model probabilities of ca. \num{0.3} and dominates over reliability.

\subsection{Skill evaluation metrics}\label{sec:skill_scores}
Metrics for evaluating classification skill using a test set with $N$ examples include the Brier score (BS), 
\begin{equation}
\text{BS} = \sum_{k=1}^{N} (p^{(k)} - y^{(k)})^2, \qquad p^{(k)} = f_\mathbf{w}(\bm{\xi}^{(k)},g),
\label{eq:BS}
\end{equation}
which is known for being strictly proper \citep{Broecker2007b}. Normalization with a reference Brier score $\text{BS}_\text{ref} = \sum_{k=1}^{N} (g - y^{(k)})^2$ of a random climatological model yields the Brier skill score (BSS)
\begin{equation}
\text{BSS} = 1-\frac{\text{BS}}{\text{BS}_\text{ref}}
\end{equation}
\citet{Murphy1973} showed that BSS can be written as the difference between resolution and reliability (\cref{sec:model_reliability}). Thus, in terms of \crefrange{eq:res}{eq:rel}, BSS is given by the area between RES and REL as functions of $p$. This is illustrated in \cref{fig:reliability_diagram} (b).

While the BSS directly acts on the probability outputs $p^{(k)}$ (\cref{eq:BS}) of the model, a large class of classification metrics require the conversion of probabilities to binary output first.
This is done by introducing a decision threshold $\tilde{p}$. If $p>\tilde{p}$, thunderstorm occurrence for the corresponding example is deemed "true", otherwise "false". In combination with the two options from the label, there are four possible outcomes for each example. They are presented as a contingency matrix in \cref{tab:contingency_matrix}.
\begin{table}[htbp]
\caption{Contingency matrix for binary classification.}
\label{tab:contingency_matrix}
\begin{tabular}{cccc}
\toprule
&& \multicolumn{2}{c}{ observed thunderstorm} \\
\midrule
&& true & false \\
\multirow{2}{*}{forecast thunderstorm} & true & hit & false alarm \\
 &false & miss & correct reject \\
\bottomrule
\end{tabular}
\end{table}

While there is an infinite number of options to combine the four possible outcomes to a single skill score, we selected the scores in this study based on their suitability for tasks with significant class imbalance. Namely, we do not wish to reward our model for correctly classifying the majority class. This amounts to dismissing scores which explicitly use correct rejects.

Given a test set and a fixed decision threshold, the probability of detection (POD) and false-alarm ratio (FAR) are defined by
\begin{align}
\text{POD} &= \frac{\text{hits}}{\text{hits}+\text{misses}}, \\
\text{FAR} &= \frac{\text{false alarms}}{\text{hits}+\text{false alarms}}.
\end{align}
Here, e.g. "hits" refers to the number of examples in the test set that qualify as "hit" according to \cref{tab:contingency_matrix}. POD is often referred to as recall in the ML literature, while $1-\text{FAR}$ is also known as precision.

Precision and recall need to be simultaneously optimized for a useful classifier. For instance, perfect recall is easily achieved by predicting the thunderstorm class at every occasion. 
For problems with class imbalance, a popular choice of combining the two scores consists of taking the harmonic mean, which yields the $F_1$-score:
\begin{equation}
F_1 = \frac{2}{\text{POD}^{-1}+(1-\text{FAR})^{-1}} = \frac{2\,\text{hits}}{2\,\text{hits}+\text{misses}+\text{false alarms}}
\label{eq:F1-score}
\end{equation}
Another option of combining the contingency matrix elements is given by the critical success index (CSI):
\begin{equation}
\text{CSI} = \frac{\text{hits}}{\text{hits}+\text{misses}+\text{false alarms}}
\label{eq:CSI-score}
\end{equation}
A modification of the CSI consists of subtracting as many hits as a model randomly classifying according to climatology would obtain. The equitable threat score (ETS) reads
\begin{equation}
\text{ETS} = \frac{\text{hits} - \text{hits by accident}}{\text{hits} - \text{hits by accident}+\text{misses}+\text{false alarms}},
\end{equation}
where the hits by accident amount to $g\times(\text{hits} + \text{false alarms})$.

\subsection{Baseline model}\label{sec:baseline_intro}
As thunderstorms are accompanied by convective precipitation, radar reflectivity constitutes a natural surrogate for thunderstorm storm occurrence in the nowcasting community \citep{Dixon1993, Wilson1998, Turner2004}.
ICON-D2-EPS outputs the column-maximal radar reflectivity (DBZ\textunderscore CMAX in \cref{tab:input_features}), which we refer to as reflectivity in what follows. In order to construct a baseline for comparison to SALAMA, we repeat training our model, but use only reflectivity as input. The architecture of the baseline model is identical to the one presented in \cref{fig:architecture} except for the input layer, which has now only a single node. Just like for SALAMA, the baseline model outputs the probability of thunderstorm occurrence (for the one ensemble member that produced the input reflectivity).

\begin{figure*}[htbp]
\centering
\includegraphics[width=0.49\textwidth]{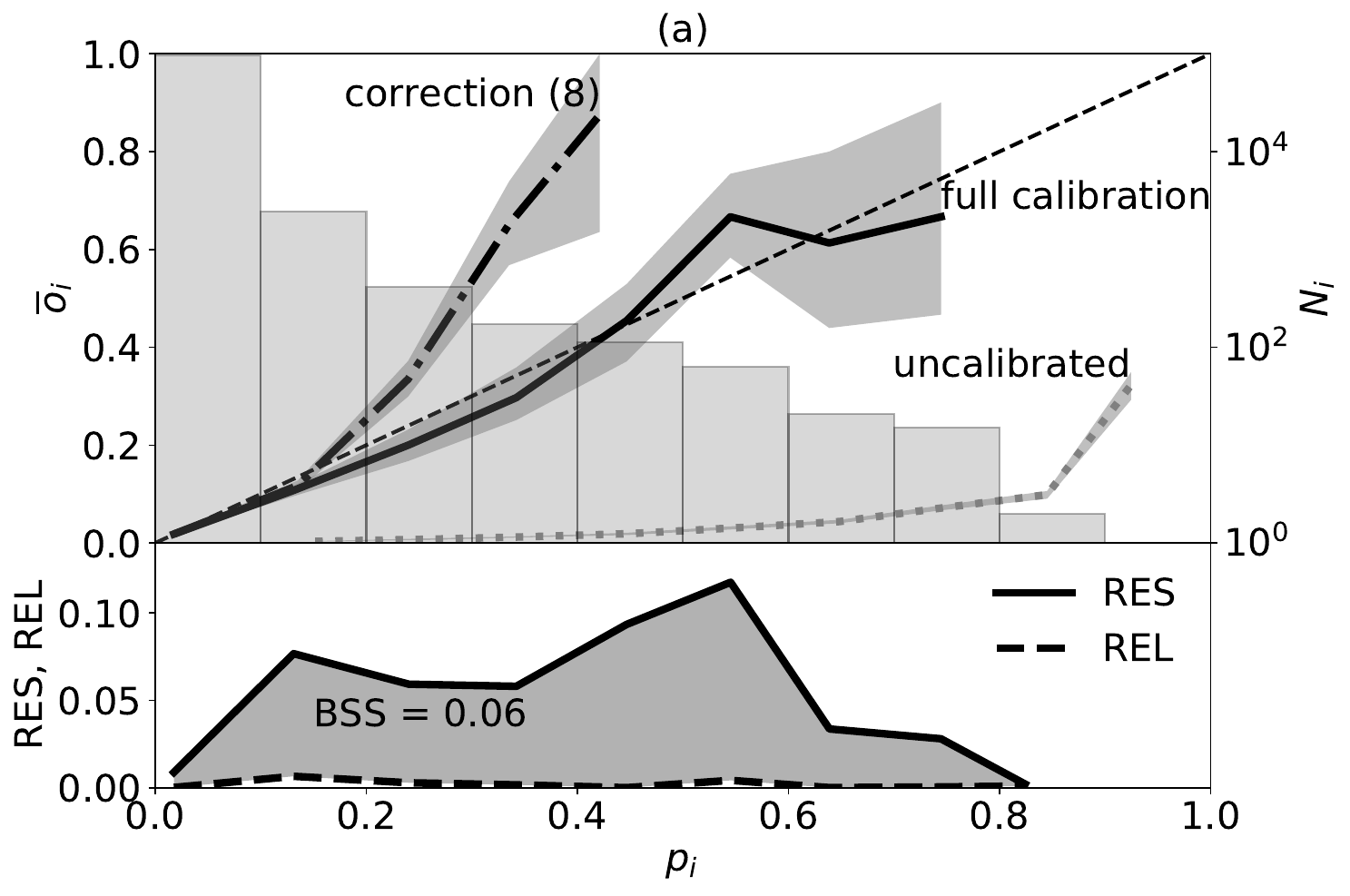}
\includegraphics[width=0.445\textwidth]{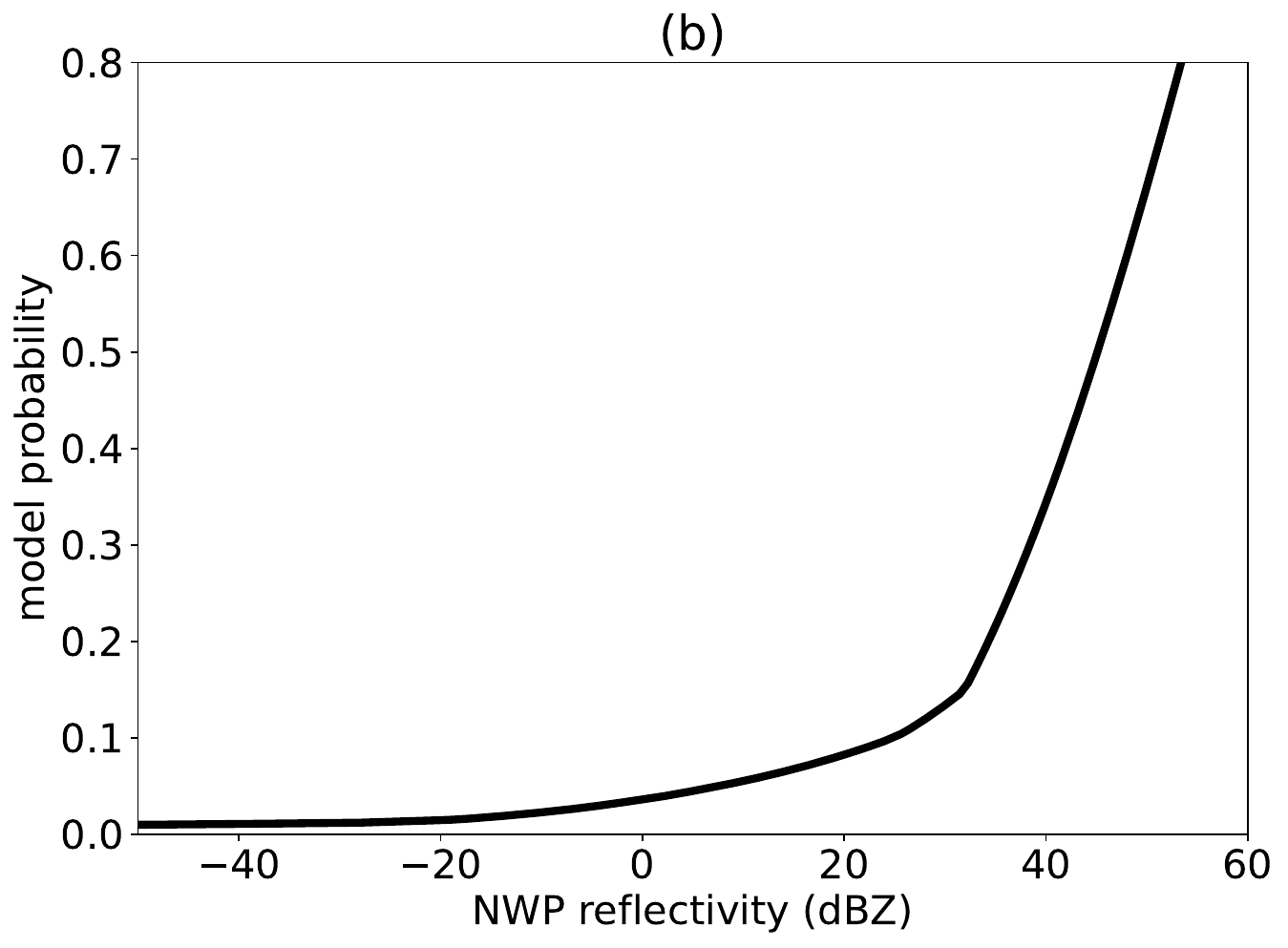}
\caption{Training of the baseline model. (a) Reliability diagram panels as in \cref{fig:reliability_diagram}, but for the baseline model. (b) Learned relationship between the baseline input field and the corresponding probability of thunderstorm occurrence.}
\label{fig:reliability_baseline}
\end{figure*}

\Cref{fig:reliability_baseline} shows the resulting reliability diagram. The light dotted line corresponds to the uncorrected calibration curve, while the dash-dotted line results from applying probability correction \eqref{eq:probability_test}. The baseline model produces well-calibrated output for small model probabilities while the model displays underconfidence above probabilities of approximately \num{0.2}. As examples with higher probabilities than \num{0.2} make up less than \SI{1}{\percent} of the examples in the test set, we assume that these examples therefore did not contribute sufficiently to the loss function, which instead favored well-calibrated small probabilities. In an effort to construct a competitive baseline model, we used the validation set to fit a linear function to the part of the dash-dotted calibration curve with probabilities higher than \num{0.15}. Then, if the output of the baseline model after application of probability correction \cref{eq:probability_test} is denoted by $p$, the calibrated output reads $C(p)$ for $p>\num{0.15}$, and $p$ otherwise. The resulting well-calibrated calibration curve is given by the solid line in the reliability diagram. The histogram and the lower panel in  \cref{fig:reliability_baseline} (a) refer to the latter calibration curve. One can see that BSS is essentially determined by the baseline resolution, just like for SALAMA (\cref{fig:reliability_diagram}). The baseline scores comparably to SALAMA in terms of reliability. On the other hand, the baseline resolution is significantly worse, which results in a lower BSS compared to SALAMA.

\Cref{fig:reliability_baseline} (b) shows the learned and calibrated relationship between NWP reflectivity and the corresponding probability of thunderstorm occurrence. The herein observed monotonously increasing relationship implies that thunderstorms become more likely as reflectivity increases. A typical threshold for defining thunderstorms in nowcasting is $\num{35}\,\text{dBZ}$ \citep{Dixon1993, Muller2003}, for which the probability of thunderstorm occurrence reads \num{0.22}.

\section{Results}\label{sec:results}
SALAMA provides a general post-processing framework for NWP ensemble forecasts. While we trained SALAMA on lead times up to two hours, we apply the same model to all lead times and all ensemble members individually, using neither the lead time nor the ensemble member index as input feature.
Working with ensemble data, our framework readily allows us to study the ensemble spread of thunderstorm occurrence. For example, if we have, for a given location, a \SI{3}{\hour}-forecast of ICON-D2-EPS at hand, it consists of 20 input feature tuples (one tuple for each ensemble member). One can now compute a thunderstorm probability according to \cref{eq:probability_test} for each member.  As we will discuss in \cref{sec:lead-dependence}, the ensemble spread of thunderstorm probability is linked to the NWP forecast uncertainty of the input features.
In the following, we compare SALAMA to the baseline model based on reflectivity (\cref{sec:baseline_intro}) and move on to investigating how the spatiotemporal thresholds of the lightning label configuration (\cref{sec:lightning}) influence the classification skill of SALAMA as a function of lead time.

\subsection{Comparison to baseline model}\label{sec:baseline_comp}
In this section, we keep the thresholds of the lightning label configuration (\cref{sec:lightning}) fixed to the particular choice $\Delta r = \SI{15}{\kilo\meter}, \Delta t = \SI{30}{\minute}$. The climatological fraction of thunderstorm examples in the test set amounts to $g=\num{0.021}$ in this configuration. The results of this section, however, do not change qualitatively if another configuration is used.

\begin{figure*}[htbp]
\centering
\includegraphics[height=0.35\textwidth]{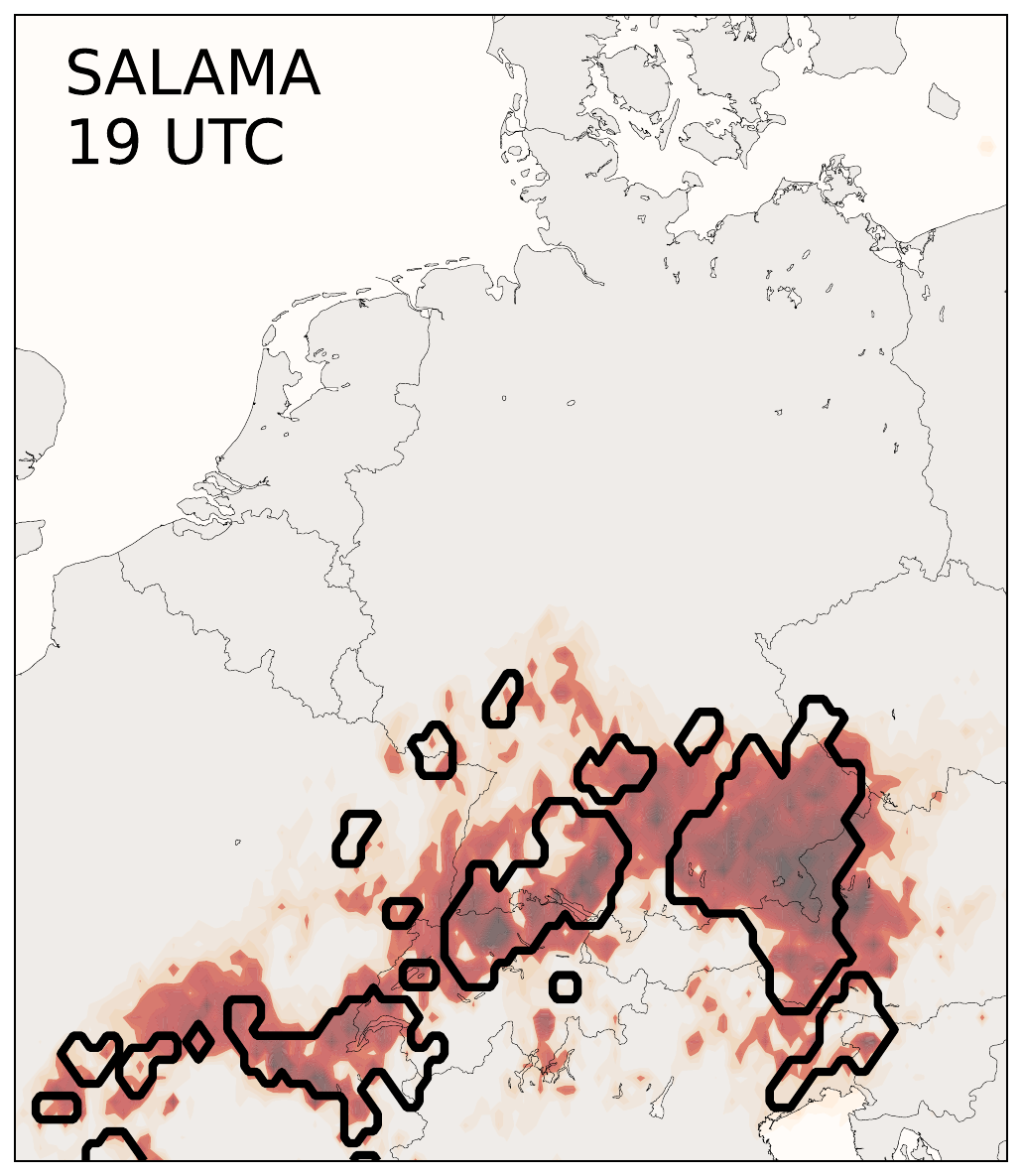}
\includegraphics[height=0.35\textwidth]{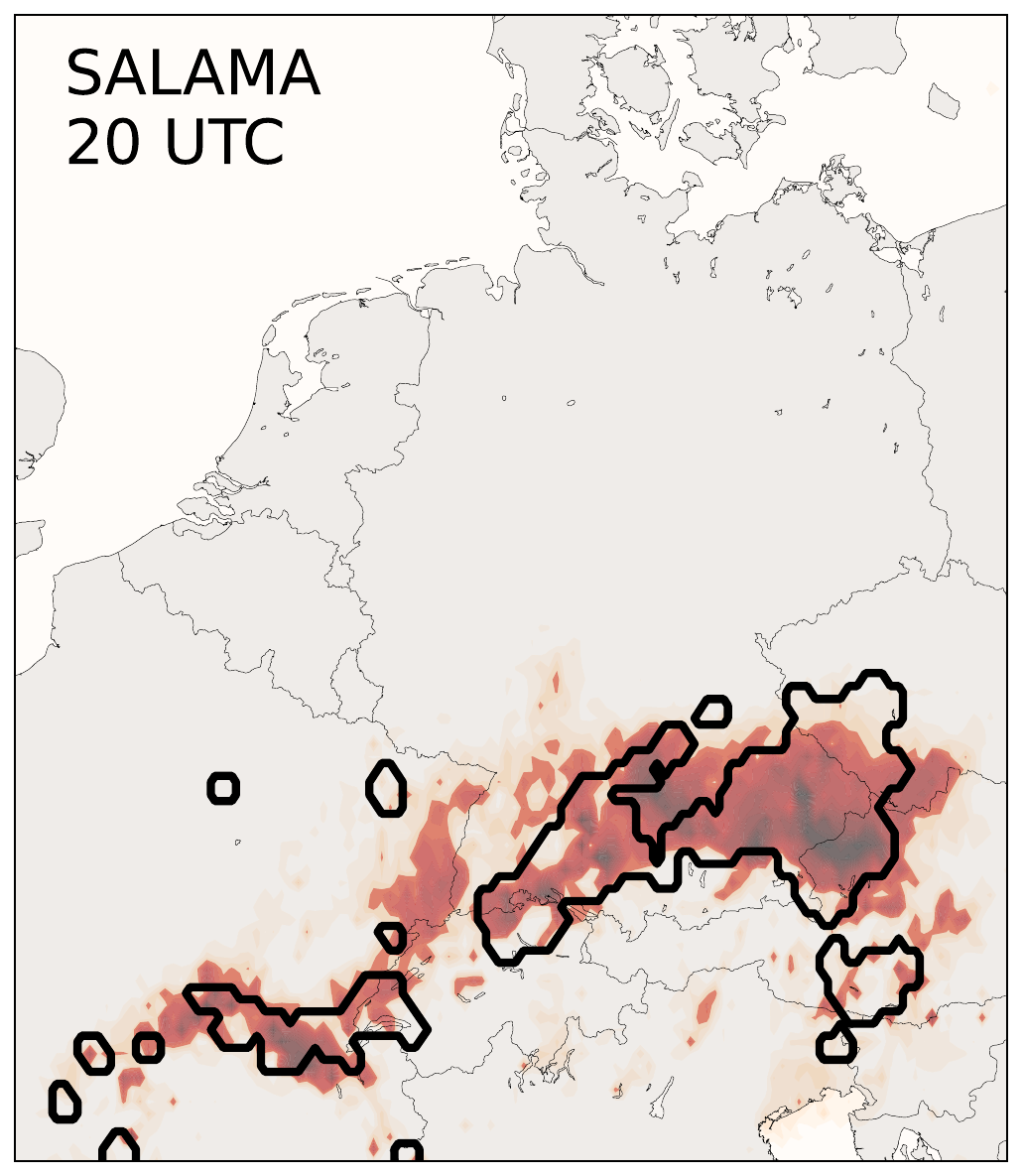}
\includegraphics[height=0.35\textwidth]{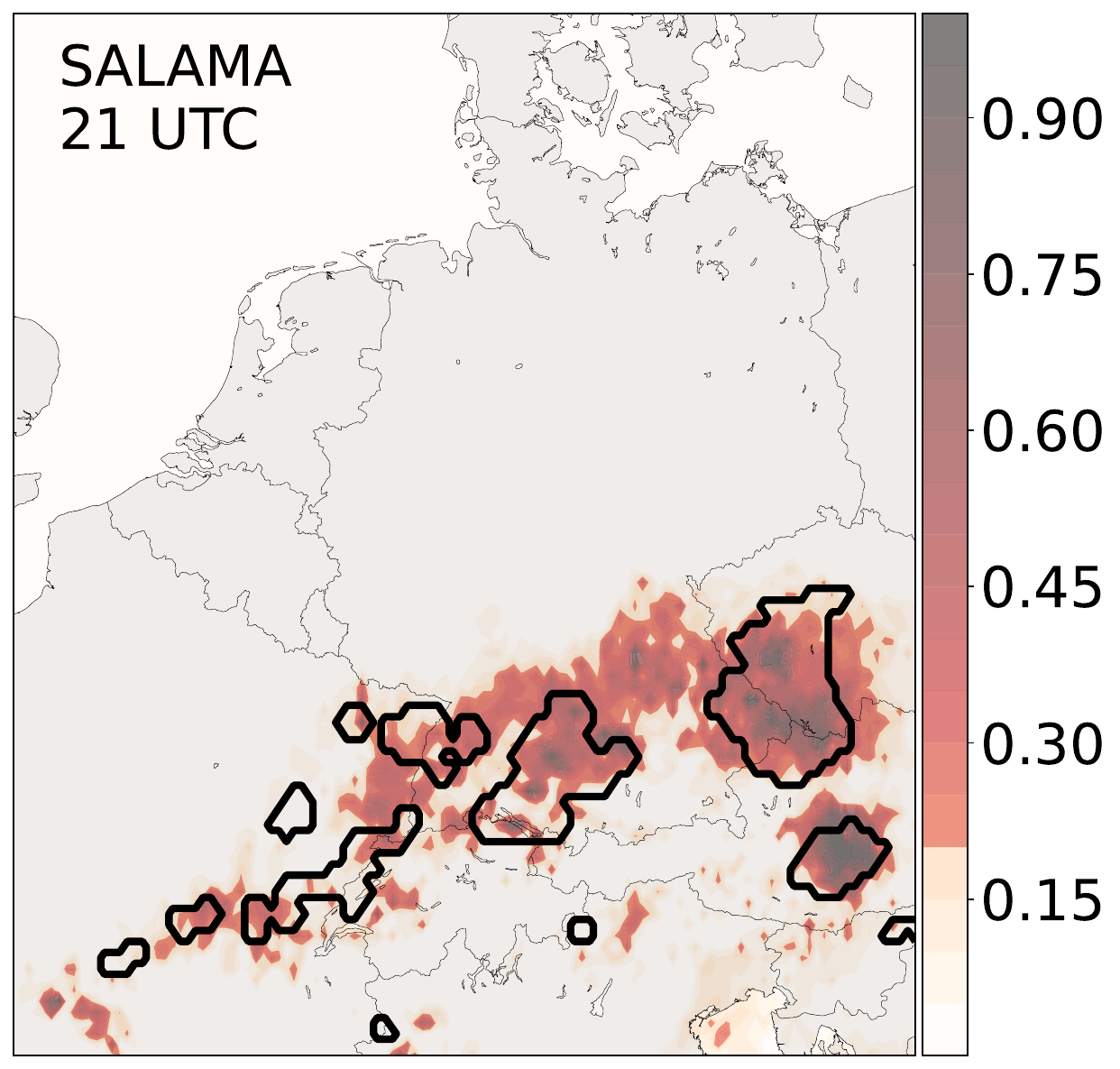}
\includegraphics[height=0.35\textwidth]{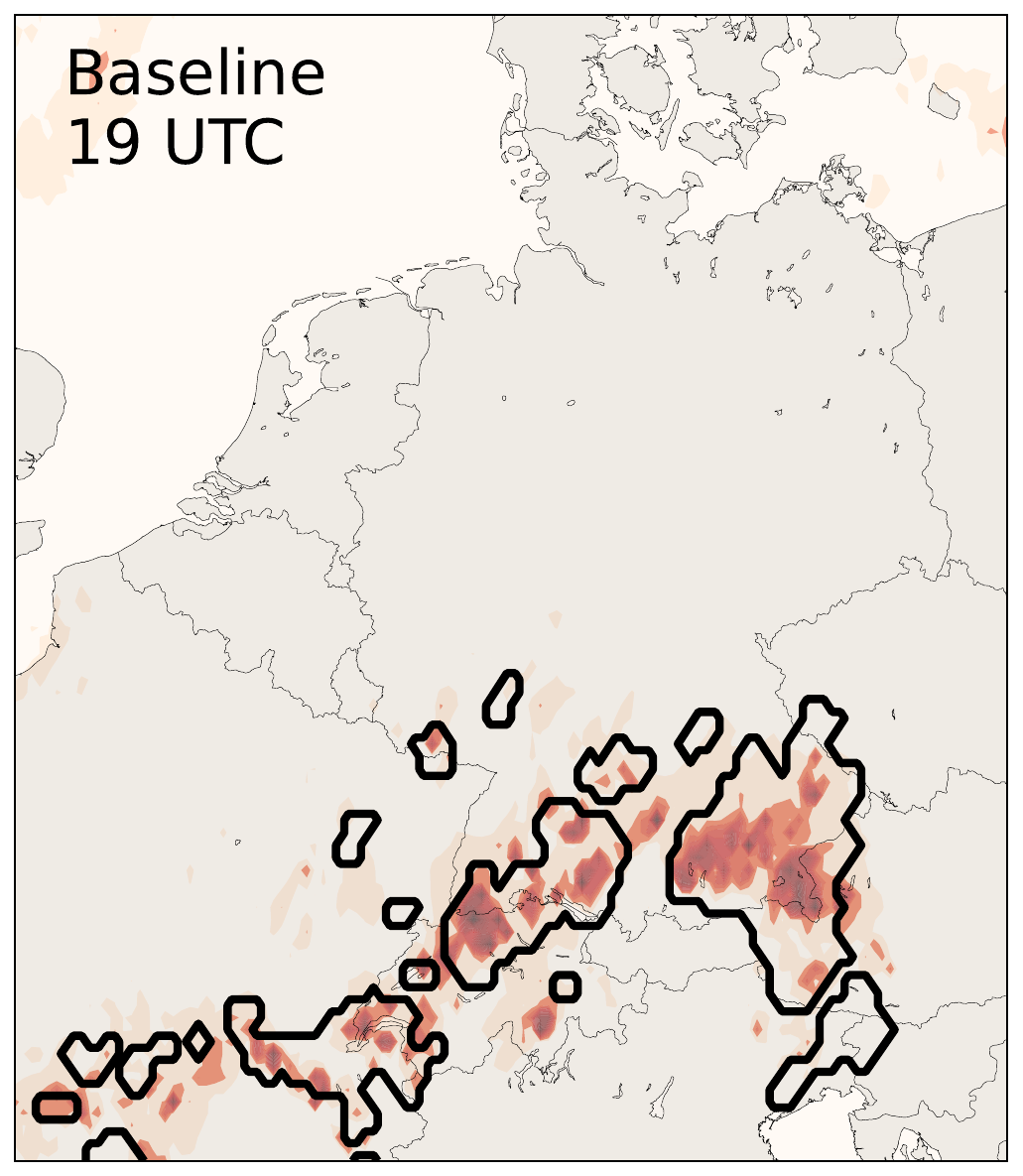}
\includegraphics[height=0.35\textwidth]{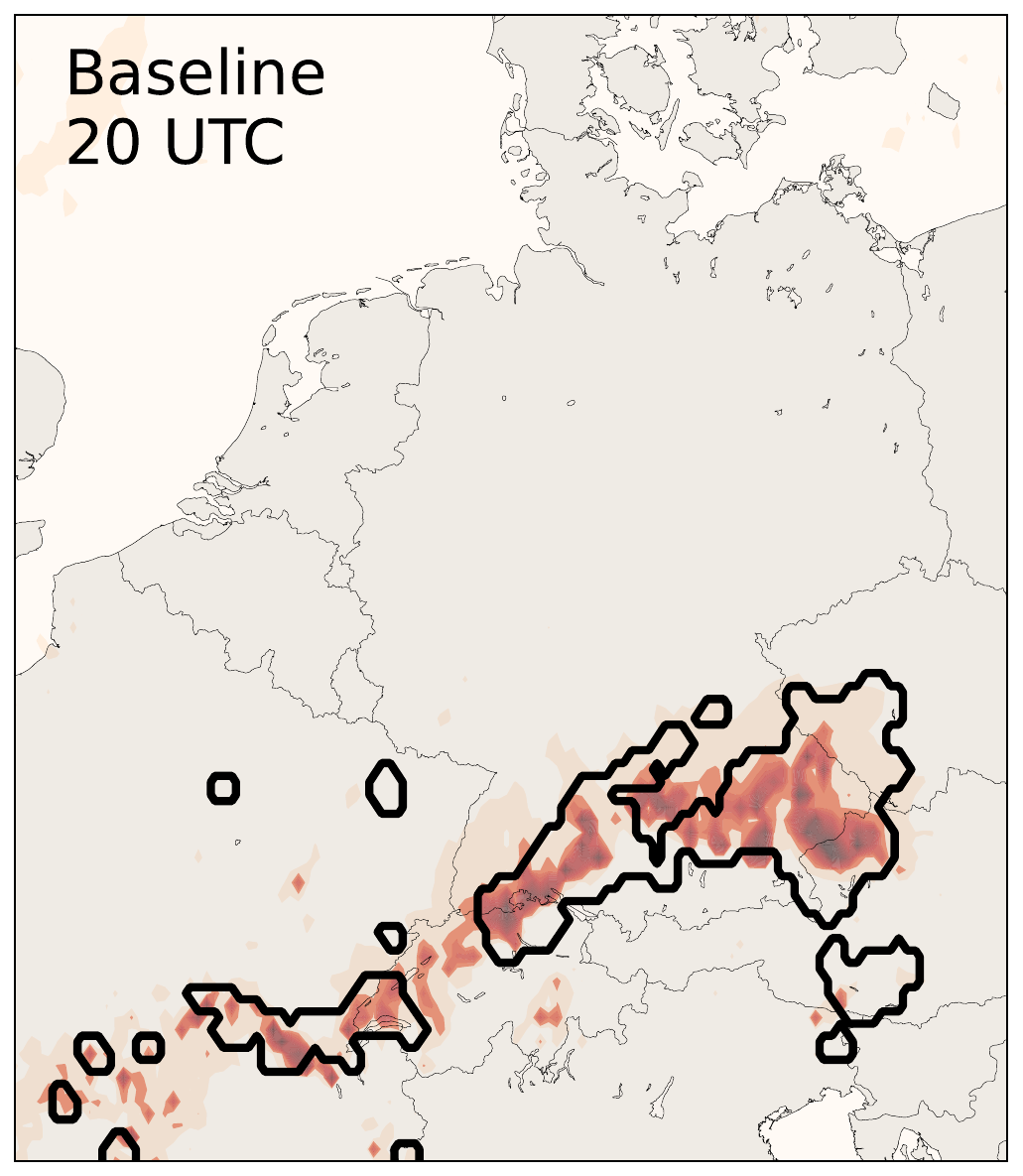}
\includegraphics[height=0.35\textwidth]{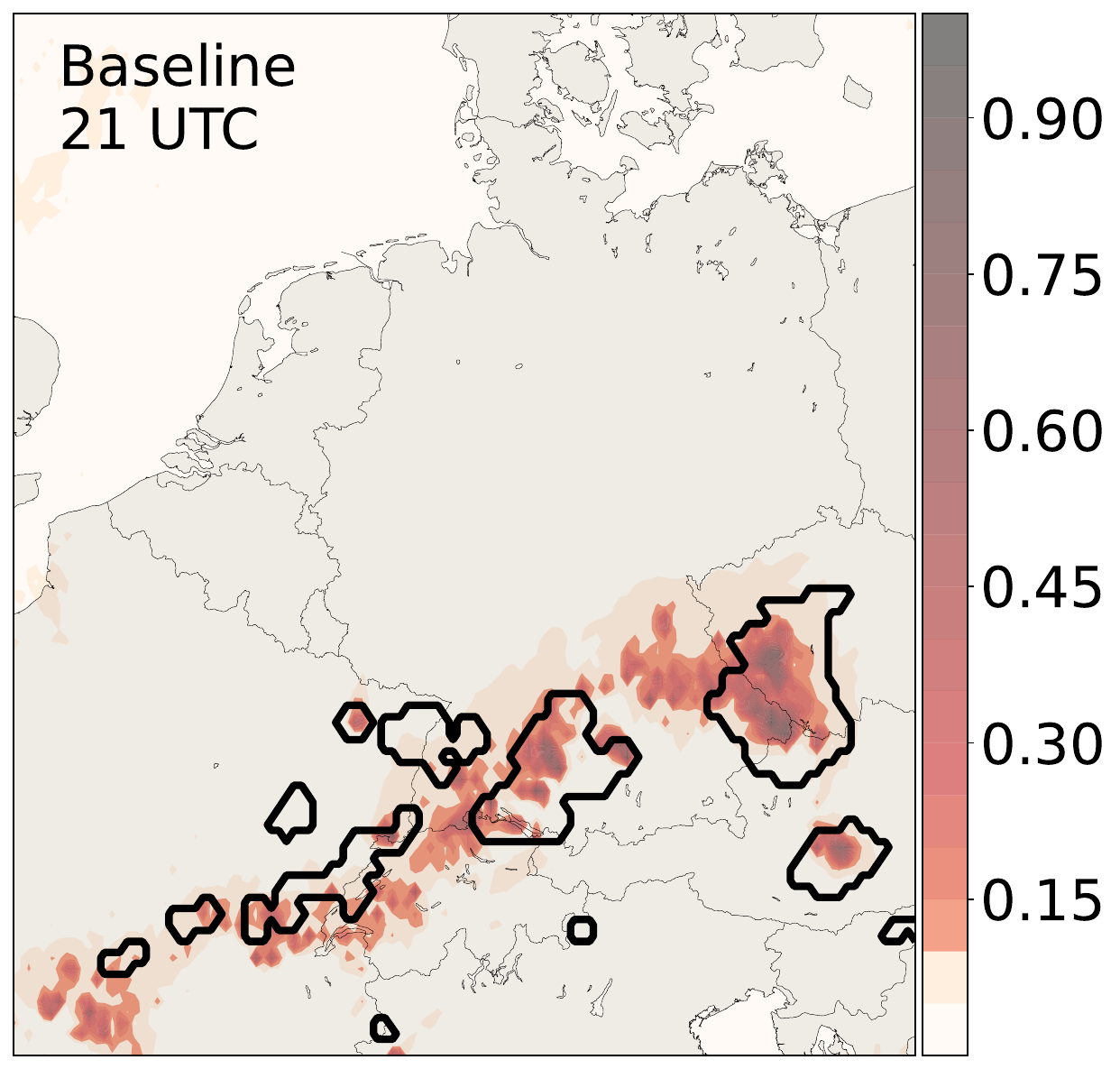}
\caption{(Color online) Probability of thunderstorm occurrence for June 23, 2021 from 19~UTC on, for SALAMA (upper row) and the baseline model (lower row). The model lead times for the three hours are \SI{1}{\hour}, \SI{2}{\hour}, and \SI{0}{\hour}, respectively. The color maps display the result for the first ensemble member of ICON-D2-EPS, while lightning labels ($\Delta r = \SI{15}{\kilo\meter}, \Delta t = \SI{30}{\minute}$, \cref{sec:lightning}) are shown as black contours.  
A jump in the color maps indicates the decision thresholds used for the evaluation of the skill scores in \cref{fig:identification_skills}.}
\label{fig:case_visualization}
\end{figure*}

As a first step, we visually compare the performance of SALAMA and the baseline model in a case study.
For this purpose, we run SALAMA  for three consecutive hours of an evening with thunderstorm occurrence over Southern Germany.  This day has not been used for the training of SALAMA. In \cref{fig:case_visualization}, we plot the probability of thunderstorm occurrence for an arbitrary member of the NWP ensemble for the entire study domain. Observed thunderstorm occurrence is given by black contours. The corresponding plots for the baseline model are added below the panels of SALAMA.
In this particular case study, SALAMA tends to detect more thunderstorm pixels than the baseline model. On the other hand,   SALAMA seems to produce more false alarms.

In order to compare the skill of SALAMA and our baseline quantitatively for the entire study period, we evaluate the skill scores introduced in \cref{sec:skill_scores}. 
We use for this purpose the test set introduced in \cref{sec:datasets}, which consists of examples of the entire summer of 2021. 
For some of the scores, we need to set a decision threshold.
As a criterion, we demand that forecasts be unbiased (average fraction of examples classified as thunderstorms is equal to the observed fraction of thunderstorm examples), yielding thresholds of \num{0.193} (SALAMA) and \num{0.119} (baseline). The thresholds are also indicated in the color bars of \cref{fig:case_visualization}. The threshold found for reflectivity corresponds to $\num{28}\,\text{dBZ}$ and is slightly below the typical literature threshold cited in \cref{sec:baseline_intro}.

The performance of SALAMA and the baseline is summarized in \cref{fig:identification_skills}. Irrespectively of the skill score under consideration, SALAMA scores better than the baseline model. The uncertainties are computed here, as well as for the subsequent evaluations, by the bootstrap resampling method introduced in \cref{sec:model_reliability}.
Note that we obtain $\text{POD}=1-\text{FAR}=F_1$ for both models. This is a result from our choice of decision threshold: Recall generally equals precision for unbiased forecasts \citep{Wilks2011}.
\begin{table}[htbp]
\caption{Scores for classification skill, evaluated on the test set, for SALAMA and the baseline model. The probability thresholds used for evaluation are chosen such that the forecast is unbiased and amount to \num{0.193} (SALAMA), \num{0.119} (baseline). Uncertainties are obtained from 200 bootstrap resamples and show the symmetric \SI{90}{\percent} confidence interval.}
\label{fig:identification_skills}
\begin{tabular}{rll}
\toprule
skill score & SALAMA & Baseline\\
\midrule
PR-AUC & \num{0.358(18)} & \num{0.141(12)}\\
BSS & \num{0.209(10)} & \num{0.063(7)} \\
POD & \num{0.403(16)} & \num{0.189(12)} \\
1-FAR & \num{0.402(17)} & \num{0.188(13)} \\
$F_1$ & \num{0.403(15)} & \num{0.189(12)} \\
CSI & \num{0.252(12)} & \num{0.104(7)} \\
ETS & \num{0.241(12)} & \num{0.093(7)} \\
\bottomrule
\end{tabular}
\end{table}

\begin{figure}[htbp]
\centering
\includegraphics[width=\columnwidth]{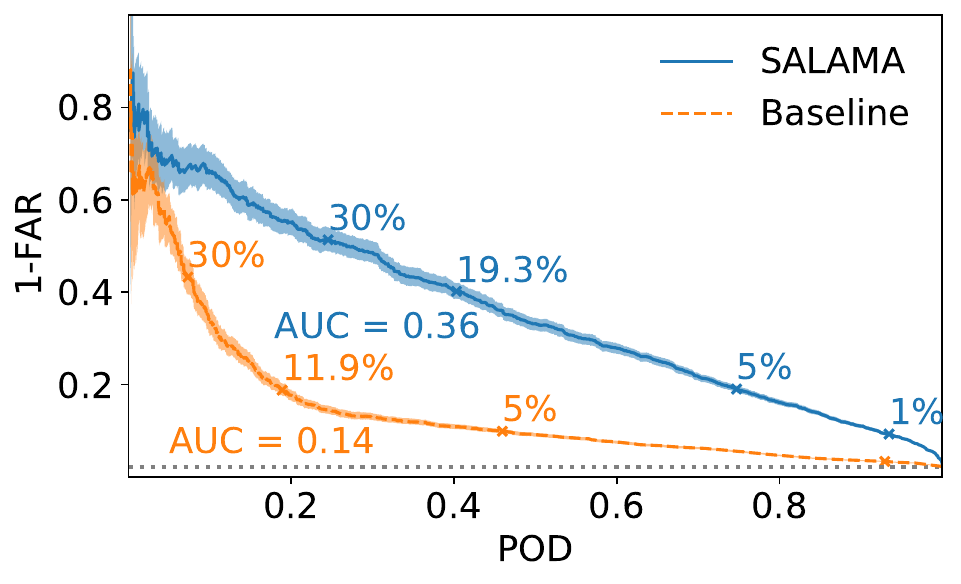}
\caption{(Color online) PR curve for SALAMA (solid) and the baseline model (dashed), evaluated on the test set. The annotations added to the curves correspond to different decision thresholds (\cref{sec:skill_scores}).
Grey dotted line denotes models with no identification skill. Uncertainties are obtained from 200 bootstrap resamples and show the symmetric \SI{90}{\percent} confidence interval.}
\label{fig:ROC_identification}
\end{figure}

Drawing $(\text{POD},1-\text{FAR})$ for different decision thresholds into one diagram, one obtains the precision-recall (PR) diagram in \cref{fig:ROC_identification}.  A random model with no skill corresponds to the dashed horizontal curve $1-\text{FAR}=g$, where $g$ denotes the climatological fraction of positive examples in the test set. Models with skill display PR curves above the horizontal line, with higher areas under the curve (AUC) indicating higher classification skill.
Both models considered in this study display higher skill than a random model following climatology would. SALAMA, however, has higher classification skill than the baseline, as can be seen from the higher AUC in the PR curve in \cref{fig:ROC_identification}. The enhanced skill of SALAMA with respect to the baseline model illustrates that a multi-parameter approach to thunderstorm forecasting is superior to employing a single input feature.


\subsection{Lead time dependence of classification skill}\label{sec:lead-dependence}
\begin{figure*}[htbp]
\centering
\includegraphics[width=\textwidth]{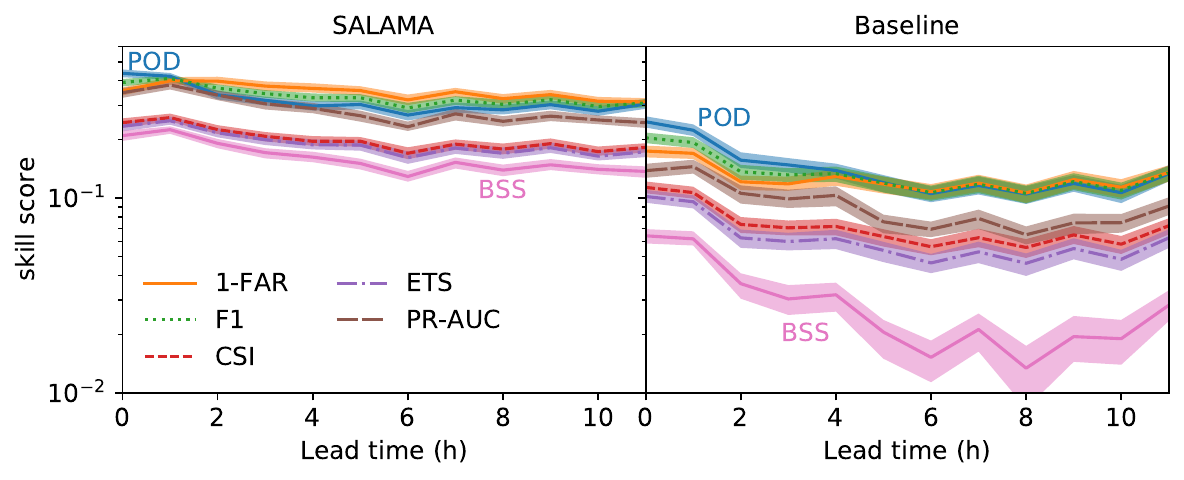}
\caption{Classification skill as a function of lead time for SALAMA (left) and the baseline model (right). The probability thresholds used for evaluation are chosen such that the forecast is unbiased and amount to \num{0.193} (SALAMA), \num{0.119} (baseline). Uncertainties are obtained from 200 bootstrap resamples and show the symmetric \SI{90}{\percent} confidence interval.}
\label{fig:scores_leadtime}
\end{figure*}

The data sets for training, testing and validation introduced in \cref{sec:datasets} and used in \cref{sec:baseline_comp} are comprised of NWP forecasts with lead times up to \SI{2}{\hour}. The reason for this choice was to train and evaluate our model in a setting of minimal NWP forecast uncertainty. On the other hand, this procedure raises the question whether the thunderstorm signature learned by the model generalizes to NWP data with longer lead times (and higher forecast uncertainty).
For this purpose, we generate test sets in which the examples come from NWP forecasts with fixed lead time. Each set contains \num{e5} examples. We use the same dates as for the test sets introduced in \cref{sec:datasets}. In \cref{fig:scores_leadtime},  we plot the SALAMA classification skill, measured in terms of the skill scores introduced in \cref{sec:skill_scores} as a function of lead time and compare it to the dependence obtained for the baseline model.  \Cref{fig:scores_leadtime} shows that, for SALAMA,  classification skill decreases approximately exponentially (note the log-scaling of the $y$-axis) for lead times longer than \SI{1}{\hour}, irrespectively of the skill score under consideration. The classification skill of SALAMA at a lead time of \SI{1}{\hour} is actually higher than at \SI{0}{\hour}, which is likely a spin-up effect from the NWP model \citep{Sun2014}.
On the other hand, SALAMA skill is systematically superior to baseline skill for all lead times. In fact, even the 11-hour-lead-time skill of SALAMA is higher than the baseline skill for any of the considered lead times.


\begin{figure}[htbp]
\centering
\includegraphics[width=\columnwidth]{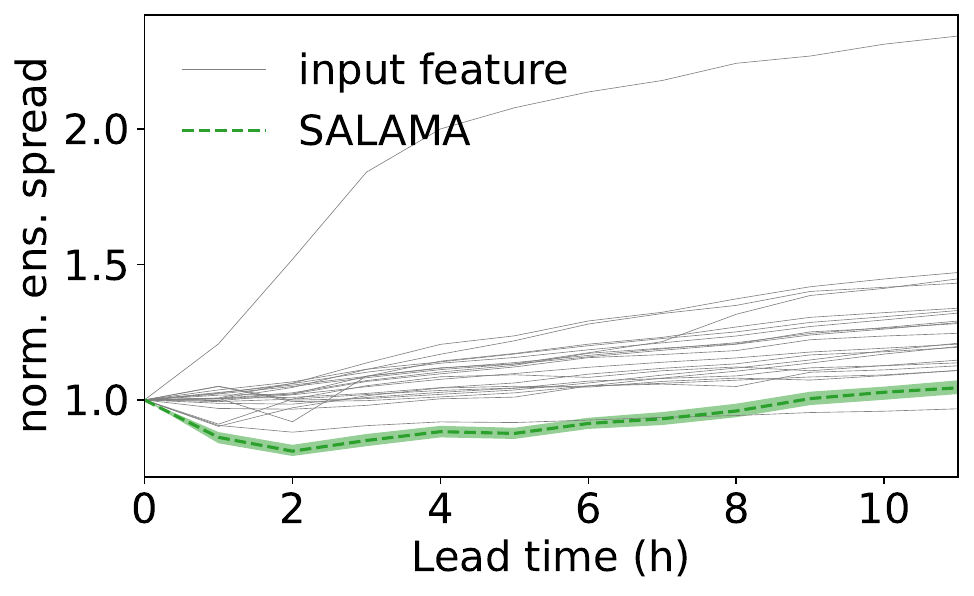}
\caption{Normalized ensemble spread (as defined in \cref{eq:ensemble_spread_avg}) of input features in comparison to spread of model thunderstorm probability as a function of lead time. Each thin solid line refers to one of the 21 input features. The thick dashed green line is associated with SALAMA. A shaded band represents the symmetric \SI{90}{\percent} confidence interval of uncertainty, estimated with 200 bootstrap resamples.}
\label{fig:variance_in-out}
\end{figure}

It is tempting to assume that the decrease in skill with lead time originates from an increasing NWP forecast uncertainty for longer lead times. We can use ensemble data to check this hypothesis. Let $q$ be either one of the 21 input features or the model thunderstorm probability, i.e. a quantity that is given for each ensemble member and each lead time. Then define the ensemble spread $\sigma'_q$ of $q$ as the ensemble standard deviation of $q$,
\begin{equation}
\sigma'_q (t_\text{lead}) = \sqrt{\langle q(t_\text{lead})^2 \rangle - \langle q(t_\text{lead}) \rangle^2},
\label{eq:ensemble_spread}
\end{equation}
where we make the dependence on the lead time $t_\text{lead}$ explicit. The brackets $\langle \cdot \rangle$ denote the average over all 20 ensemble members. Denote by $\overline{\sigma'_q}(t_\text{lead})$ the expression obtained by performing the average of $\sigma'_q$ over the entire study region and all times associated with the test set.
Lastly, we define the normalized ensemble spread of $q$,
\begin{equation}
\sigma_q (t_\text{lead}) = \frac{\overline{\sigma'_q}(t_\text{lead})}{\overline{\sigma'_q}(\SI{0}{\hour})},
\label{eq:ensemble_spread_avg}
\end{equation}
as a function of lead time. It quantifies ensemble spread in such a way that different input features can be directly compared to each other. In \cref{fig:variance_in-out}, the normalized ensemble spread of each of the 21 input features is shown as thin solid lines and the corresponding curve for the model output of SALAMA is drawn in thick and dashed. One can see that the ensemble spread does indeed increase with lead time for most input features, the increase being approximately linear. The ensemble spread of the SALAMA output increases in line with the majority of the input features and with a similar slope. This suggests that the decrease in classification skill  observed in \cref{fig:ROC_identification} is solely due to the increasing variance in the simulation data.

\subsection{Effect of the label size}\label{sec:label_role}
So far, the temporal and spatial thresholds of the label configuration have been fixed to $\Delta r = \SI{15}{\kilo\meter}$ and $\Delta t = \SI{30}{\minute}$ (henceforth referred to as default configuration). In this section, we study how varying the spatiotemporal thresholds affects the classification skill of SALAMA.

\begin{figure*}[htbp]
\centering
\includegraphics[width=0.33\textwidth]{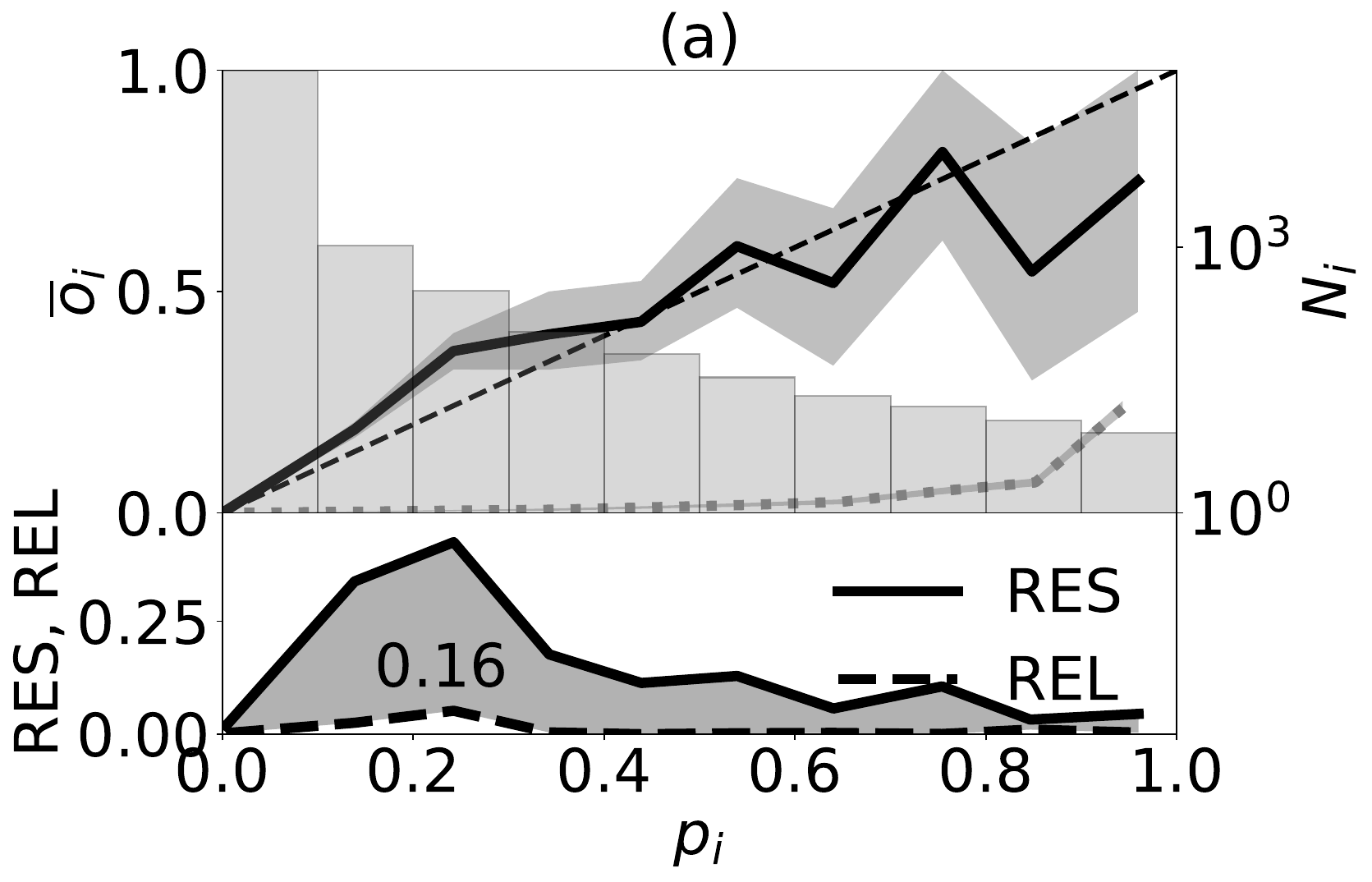}
\includegraphics[width=0.33\textwidth]{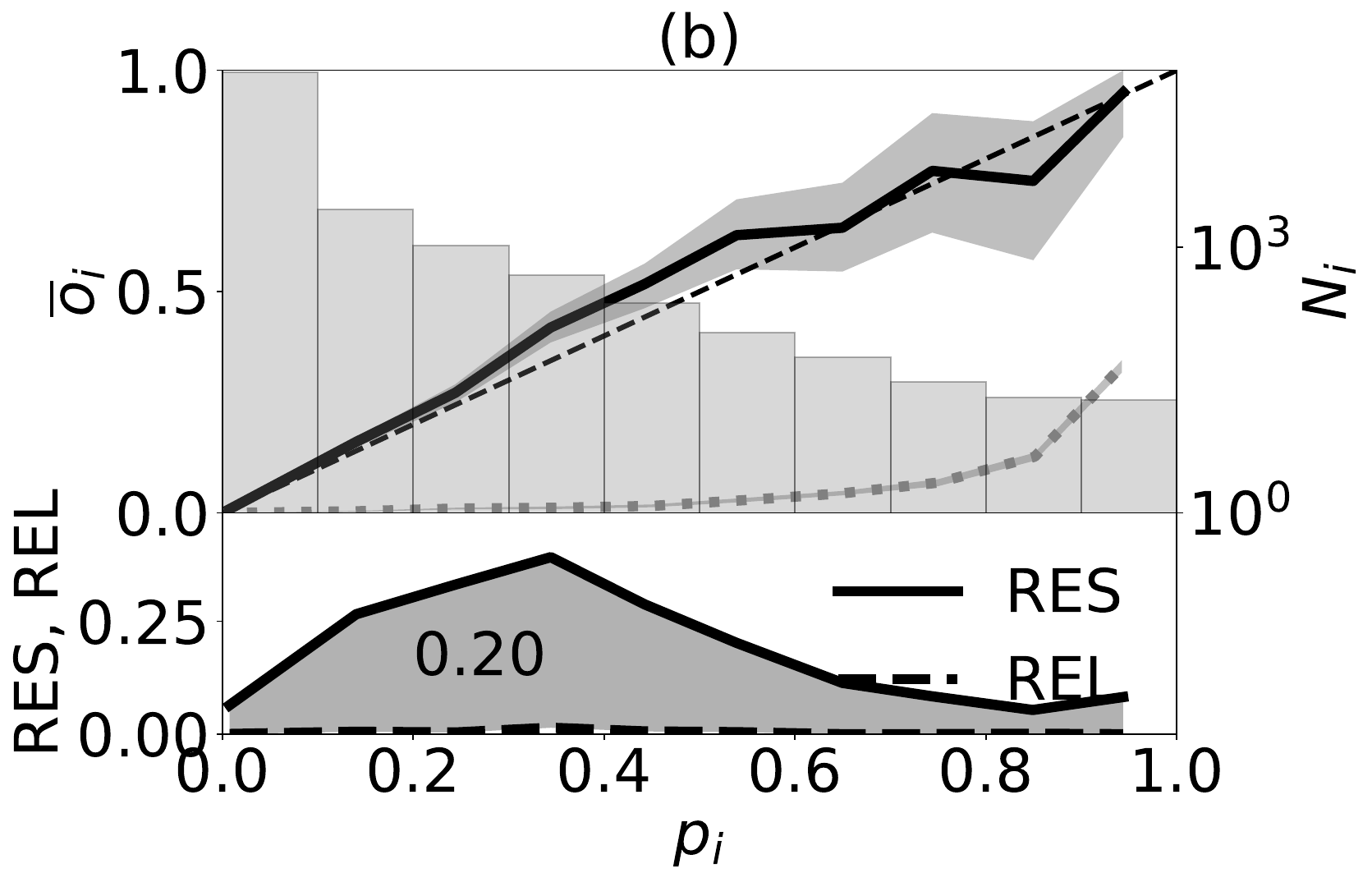}
\includegraphics[width=0.33\textwidth]{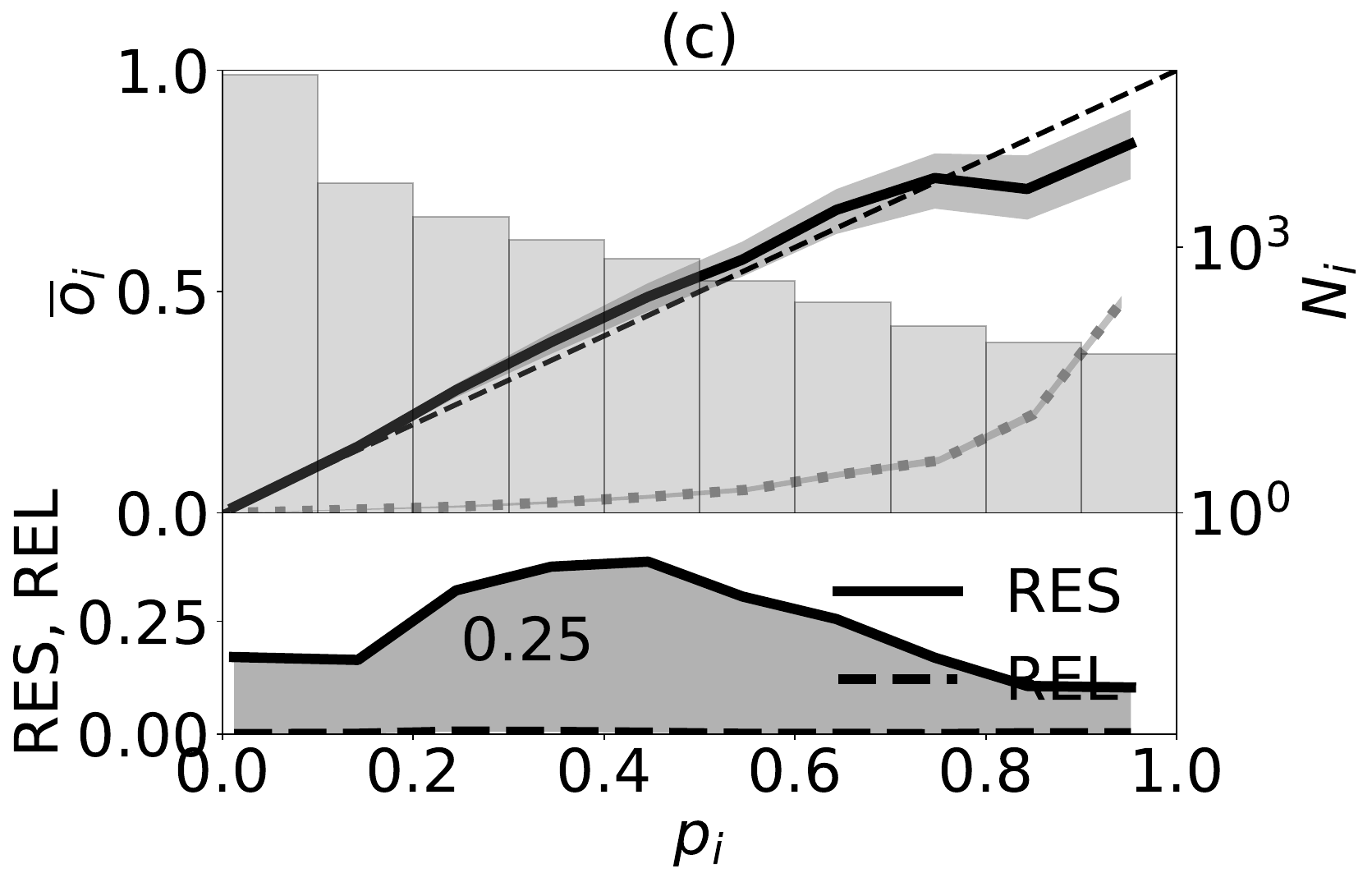}
\caption{Reliability diagrams as in \cref{fig:reliability_diagram}, but with label configurations (a) $\Delta t = \SI{15}{min}, \Delta r = \SI{9}{\kilo\meter}$ ($s = \SI{14}{\kilo\meter}$), (b) $\Delta t = \SI{10}{min}, \Delta r = \SI{21}{\kilo\meter}$ ($s = \SI{24}{\kilo\meter}$), (c) $\Delta t = \SI{40}{min}, \Delta r = \SI{24}{\kilo\meter}$ ($s = \SI{36}{\kilo\meter}$). The spatial scale $s$ is introduced in \cref{eq:forecast_scale}.}
\label{fig:more_reliability_diagrams}
\end{figure*}

As a first step, we compute reliability diagrams for different label configurations. In panel (a) of \cref{fig:more_reliability_diagrams}, we study a configuration with smaller thresholds than for the configuration studied so far. Panel (b) displays a configuration with reduced $\Delta t$ and increased $\Delta r$. In panel (c), both thresholds are increased with respect to the default configuration. The exact choice of $\Delta t$ and $\Delta r$ for the three panels is somewhat arbitrary but still allows for qualitative insight:
Irrespectively of the configuration, forecasts are well-calibrated for small and large model probabilities. In addition, model skill, quantified in terms of BSS, increases from left to right. The diagrams show that the increase in BSS is mainly due to enhanced contribution to resolution from probabilities larger than \num{0.3}, though a reliability improvement from probabilities around \num{0.2} adds to the increase in BSS as well.

\begin{figure}[htbp]
\includegraphics[width=\columnwidth]{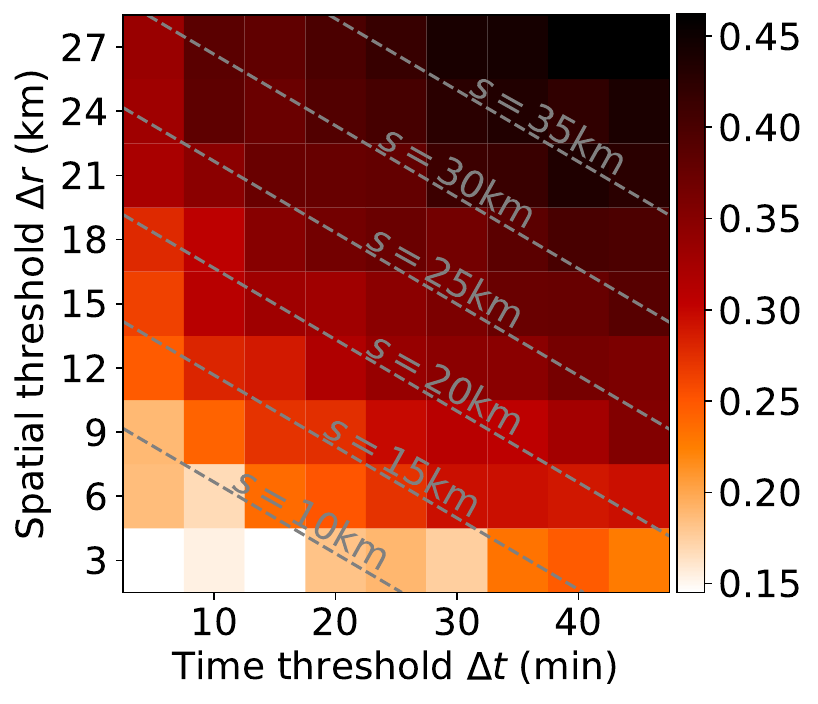}
\caption{(Color online) Classification skill of SALAMA, expressed in terms of the area under the PR curve, as a function of the label configuration (\cref{sec:lightning}).  The slope of the dashed lines is chosen such that classification skill is approximately constant along the lines. Each line corresponds to a specific spatial scale $s$ (\cref{eq:forecast_scale}). }
\label{fig:mean_wind}
\end{figure}

As we have seen in \cref{sec:lead-dependence} that the qualitative lead time dependence of SALAMA skill does not depend on the skill score, we consider from now on only PR-AUC for further investigations.  We start by computing PR-AUC for several label configurations, which is shown in \cref{fig:mean_wind}. The color pattern in the figure suggests that the two thresholds are not independent variables of classification skill. Instead, one can find a parameter $c$ with the units of a velocity such that classification skill is nearly constant along lines
\begin{equation}
   s = \Delta r + c \Delta t = \text{const}.
   \label{eq:forecast_scale}
\end{equation}
Indeed, $s$ corresponds to a spatial resolution scale; it determines the minimal spatial accuracy that can be expected from a model trained with a given label configuration. We expect the parameter $c$ to roughly quantify the speed at which regions of thunderstorm occurrence are advected in the atmosphere. A fit to the data provides $c = \SI{5.6(3)}{\meter\per\second}$, which is similar to typical low- to mid-tropospheric wind speeds in Central Europe. 
We can now motivate the spatiotemporal thresholds for the reliability diagram in the middle panel (\cref{fig:more_reliability_diagrams}): they have been chosen such that $s$ takes on the same value as the default configuration.

Lines of constant $s$ appear as dashed lines in \cref{fig:mean_wind} and indicate that classification skill increases with $s$. This is in line with the displayed observation of increased BSS in the reliability diagrams. This is also consistent with the work of \cite{Roberts2008}, which investigates the spatial variation of precipitation forecast skill. Note that sample climatology $g$ increases with $s$ as well. In fact, it amounts to $g=\num{1.7e-3}$ in the lower left pixel of \cref{fig:mean_wind}, and to $g=\num{4.6e-2}$ in the upper right corner. Since a random model with no skill has PR-AUC $= g$ (\cref{sec:baseline_comp}), the increase in skill as a function of $s$ is to a slight extent also due to the increase in $g$. 

\begin{figure}[htbp]
\includegraphics[width=\columnwidth]{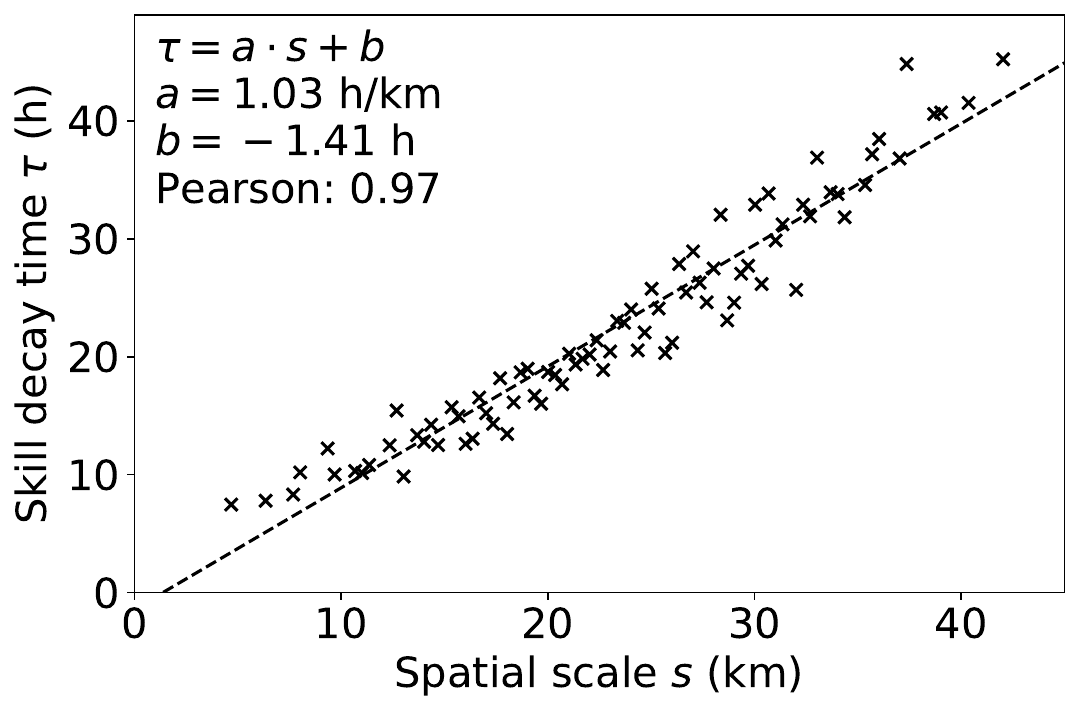}
\caption{Decay time of classification skill (quantified by the area under the PR curve) as a function of the spatial scale. Each data point corresponds to one label configuration in \cref{fig:mean_wind}. The parameters of a linear fit are also shown, as well as the Pearson coefficient of correlation.}
\label{fig:skill_decay}
\end{figure}

Next, we investigate how the decrease of classification skill with lead time depends on the spatial scale. Motivated by the observed decay of classification skill with lead time (\cref{sec:lead-dependence}), we fit an exponential function $\exp(-t_\text{lead}/\tau)$ to the lead time dependence of classification skill (measured again by the area under the PR curve). The skill decay time $\tau$ then provides a characteristic time scale for the decrease of classification skill.  For each label configuration in \cref{fig:mean_wind}, we compute the corresponding spatial scale as well as $\tau$. In \cref{fig:skill_decay}, we present a scatter plot of $\tau$ and $s$. The figure shows a tight positive linear correlation between the two quantities, which means that classification skill decreases more slowly for coarser label configurations. This is in agreement with the  anticipation \citep{Lorenz1969} that resolving smaller scales in NWP models is associated with a more rapid growth of forecast errors. Our finding is complementary to convection studies involving a scale-dependent skill score \citep{Roberts2008}, and high-resolution simulations \citep{Selz2015}.

\section{Conclusion and perspectives}\label{sec:conclusion}

Addressing the need for accurate thunderstorm forecasting and leveraging advances in high-resolution NWP and ML, we have presented SALAMA, a feedforward neural network model that identifies thunderstorm occurrence in NWP forecasts up to \SI{11}{\hour} in advance in a pixel-wise manner. The inference of the probability of thunderstorm occurrence is based on input parameters that are physically related to thunderstorm activity and do not explicitly feature information on location, time or forecast range. This gives reason to expect that the signature learned by the model generalizes to thunderstorms outside the study region of this work and remains valid in a changing climate. In addition, the availability of all input features in real-time makes SALAMA readily available for operational use.

We have addressed the technical challenge caused by the rarity of thunderstorms and the corresponding small fraction of positive examples by increasing this fraction during training and analytically accounting for the increase when testing. This approach has allowed us to ensure reasonable reliability without calibration fits. 
Furthermore, we have proposed a novel visualization of reliability and resolution as a function of bin-wise model probability. The visualization arguably proves useful for evaluating how examples with a certain model probability contribute to classification skill.

Working with ensemble data, we have studied how the NWP forecast uncertainty depends on the lead time of the forecast and have related it to the classification skill decrease of SALAMA. This has suggested that the decrease in skill is the result of an increasing uncertainty in the input feature forecasting. 

During the training process, we have systematically varied the spatiotemporal criteria by which we associate lightning observations with NWP data. This has allowed us to test SALAMA with different spatial scales and to estimate the order of magnitude of the speed at which thunderstorms are advected in the atmosphere. We have shown that classification skill increases with the spatial scale of the forecast and is higher than for a baseline model based on NWP reflectivity alone. Furthermore, we have found that the decay time of classification skill is proportional to the spatial scale. In combination with the result that the SALAMA classification skill is correlated with the NWP forecast uncertainty, our findings have indicated that resolving thunderstorms at smaller scales reduces the predictability of thunderstorm occurrence.

In a future work, it is useful to check the universality of the thunderstorm signature learned by SALAMA, e.g. by testing it on data outside of Central Europe or for a different time period than the summer of 2021. Moreover, one may explore whether classification skill can be improved by shifting from a pixel-wise consideration of input features to taking their spatiotemporal structure into account as well.

\section{Author contributions}
\textbf{K. Vahid Yousefnia:} Methodology; software; investigation; resources; visualization; writing - original draft; writing - review and editing.
\textbf{T. Bölle:} Conceptualization; writing - review and editing.
\textbf{I. Zöbisch:} Conceptualization; resources; writing - review and editing.
\textbf{T. Gerz:} Writing - review and editing; supervision.

\section{Data availability}
The Python code for SALAMA will be made available upon reasonable request.

\section{Acknowledgements}
We thank George Craig and Tobias Selz for helpful discussions. This work was funded through the internal project DIAL of the German Aerospace Center (DLR). We gratefully acknowledge the computational and data resources provided through the joint high-performance data analytics (HPDA) project "terrabyte" of the DLR and the Leibniz Supercomputing Center (LRZ). The authors declare that there are no conflicts of interest to disclose.

\printendnotes

\bibliography{main.bbl}



\end{document}